\documentclass[11pt]{article}
\usepackage{amssymb}

 \topmargin - 0,5  cm
\newcommand{\dem}{\noindent{\bf Proof. }}

\newcommand{\n}{\noindent}

\newtheorem{theor}{Theorem}[section]

\newtheorem{prop}[theor]{Proposition}
\newtheorem{coro}[theor]{Corollary}
\newtheorem{lema}[theor]{Lemma}
\newtheorem{rema}{Remark}

\newtheorem{defi}{Definition}[section]

\newcommand{\be}{\begin{equation}}
\newcommand{\ee}{\end{equation}}
\newcommand{\bea}{\begin{eqnarray}}
\newcommand{\eea}{\end{eqnarray}}
\newcommand{\beas}{\begin{eqnarray*}}
\newcommand{\eeas}{\end{eqnarray*}}
\newcommand{\ben}{\begin{enumerate}}
\newcommand{\een}{\end{enumerate}}

\newcommand{\bdefi}{\begin{defi}}
\newcommand{\edefi}{\end{defi}}
\newcommand{\bsat}{\begin{theorem}}
\newcommand{\esat}{\end{theorem}}
\newcommand{\bprop}{\begin{prop}}
\newcommand{\eprop}{\end{prop}}
\newcommand{\elem}{\end{lemma}}
\newcommand{\brem}{\begin{rem}}
\newcommand{\erem}{\end{rem}}

\newcommand{\ra}{\rightarrow}

\let \frak= \frak
\let \x = \backslash

\let \n = \noindent

 \let \x = \backslash

\begin{document}
\begin{titlepage}

\vspace*{1.5cm}

\begin{center}
{\huge Generalized double extension and descriptions of quadratic Lie superalgebras }
 
\vspace*{1.25cm}

{\large Ignacio Bajo}$^{\mbox{a}}$, {\large Sa\"{\i}d Benayadi}$^{\mbox{b}}$, 
{\large Martin Bordemann}$^{\mbox{c}}$
\vspace*{0.5cm}

$^{\mbox{a}}$
{Depto. Matem\'{a}tica Aplicada II, E.T.S.I.
Telecomunicaci\'{o}n, Universidad de Vigo, Campus Marcosende, 36280 Vigo, Spain.}
{\it E-mail:} ibajo@dma.uvigo.es

$^{\mbox{b}}$
{D\'{e}partement de  Math\'{e}matique, Laboratoire LMAM, CNRS UMR 7122,
Universit\'{e}  Paul Verlaine-Metz,
Ile du saulcy,   57045 Metz cedex 1, France. }
{\it E-mail:} benayadi@univ-metz.fr

$^{\mbox{c}}$
{Laboratoire  de  Math\'{e}matiques,  Universit\'{e} de Haute Alsace,
4, rue des Fr\`{e}res Lumi\`{e}re, 68093 Mulhouse, France. }
{\it E-mail:} Martin.Bordemann@uha.fr
\end{center}

\vspace*{2cm}

\begin{center}
{\large{\bf Abstract}}
\end{center}

A Lie superalgebra endowed with a supersymmetric, even, non-degenerate, invariant bilinear form is called a quadratic Lie superalgebra. In this paper we give an inductive description of quadratic Lie superalgebras in terms of generalized double extensions; more precisely, we prove that every quadratic Lie superalgebra may be constructed by sucessive orthogonal sums and/or generalized double extensions. In particular,  we prove that all solvable quadratic Lie superalgebras are obtained by a sequence of generalized double extensions by one-dimensional Lie superalgebras. We also prove that every solvable quadratic Lie superalgebra is isometric to either a $T^*$-extension of certain Lie superalgebra or to an ideal of codimension one of a $T^*$-extension.

\vspace*{1cm}

\n {\it Keywords:} Simple Lie superalgebras, quadratic Lie superalgebra,  double extension,
$T^*-$extension,  generalized double extension, cohomology of Lie superalgebras

\vspace*{0.5cm}

\n {\it MSC:} 17B05, 17B20, 17B30, 17B40.

\vspace*{2cm}

\n {\it Corresponding Author:} Sa\"{\i}d Benayadi,
{\it e-mail:} benayadi@math.univ-metz.fr
\end{titlepage}
\section*{Introduction}

Let $\mathfrak{g}=\mathfrak{g}_{\bar 0}\oplus\mathfrak{g}_{\bar 1}$ a
finite dimensional Lie superalgebra
over a commutative algebraically closed field $\mathbb K$ of characteristic zero.
An invariant scalar product $B$ on $\mathfrak{g}$ is an even,  supersymmetric,
non-degenerate
bilinear form on $\mathfrak{g}$ which is also invariant, this is to say, the equality
$B([X,Y],Z)=B(X,[Y,Z])$ holds for every $X,Y,Z\in\mathfrak{g}$. In this case, we say
that $(\mathfrak{g},B)$ is a quadratic (or orthogonal or metrised) Lie superalgebra.
The basic classical Lie superalgebras (\cite{K1}, \cite{S}) and semisimple
Lie algebras endowed with the Killing form are, therefore, examples of
quadratic Lie superalgebras, but there are many other Lie superalgebras
which admit invariant scalar products even when the Killing form is identically zero.

Quadratic Lie algebras have been widely studied since they appear in connection
with many problems derived from Geometry, Physics and other disciplines. For instance,
it is well-known that the Lie algebra of a Lie group endowed with bi-invariant
pseudo-Riemannian metric turns out to be quadratic or --in terms of theoretical
physics-- that the Sugawara construction
is possible for these Lie algebras \cite{F-S2}, which shows its relevance
in Conformal Field Theory. In \cite{M-R1} Medina and Revoy introduced the
notion of double extension which results as the combination of a central
extension and a semidirect product.  This concept allowed them to give a certain
inductive description of quadratic Lie algebras. More precisely, they
proved that every quadratic Lie algebra may be constructed as a direct sum
of irreducible ones, and the latter by a sequence of double extensions. A different
construction, namely the $T^*$-extension, was given in \cite{borde}
to describe all solvable quadratic Lie algebras (actually, the study in \cite{borde}
is made not only in the case of Lie algebras, but for more general classes of
algebras).
The $T^*$-extension is essentially based on a generalized semidirect product. A different approach towards
an inductive classification of quadratic Lie algebras has been recently given by  Kath and Olbrich in \cite{K-O}.

The first attempt to use the technique of double extensions to describe quadratic
Lie superalgebras was done in \cite{BB}. In this work the notion
of double extension is defined for superalgebras, and it is shown that every
irreducible quadratic Lie superalgebra in which the centre intersects the even
part in a nontrivial way is a double extension. However, such a condition is rather
restrictive and
there do exist quadratic Lie superalgebras whose centre is contained in the
odd part. One can even find nilpotent examples (see section 4 below) in which
such a phenomenon occurs. Therefore, in order to give an inductive description
of all quadratic
Lie superalgebras there seems to be a need to
generalize the concept of double extension.
In this paper we shall introduce the notion of
{\it generalized double extension} which is at the same time a generalization of
the classical double extension and of the $T^*$-extension.

The paper is organized in 5 sections. The first one just recalls
the basic definitions and preliminaries. In section 2 we introduce the
concept of generalized double extension of quadratic Lie superalgebras
and show that classical double extensions and $T^*$-extensions may be
seen as particular cases. We then obtain a similar result to the one
obtained by Medina and Revoy in the Lie algebra case; this is done in
the third section in which we explicitly prove that every quadratic
Lie algebra may be constructed by sucessive orthogonal sums and/or
generalized double extensions. The particular case of solvable Lie
superalgebras is studied in sections 4 and 5. We first see that every
solvable quadratic Lie superalgebra may be obtained by a sequence of
generalized double extensions by one-dimensional superalgebra and then,
in Section 5, we give a description of such solvable quadratic superalgebras
via the $T^*$-extension. We show that, in contrast to the case of the
classical double extension, the $T^*$-extension allows us to describe all solvable cases.

We are indebted to M. Duflo for providing the proof of the central Theorem \ref{DL},
which allowed us to extend a preliminary result concerning $T^*$-extensions which
we had obtained (see \cite{BBBarXiv}) to the general solvable case, and also for the second example of
section 4. We also want to acknowledge the University of Vigo, the University of
Metz, and  the Graduiertenkolleg `Partielle Differentialgleichungen' of the University
of Freiburg for the reserch stays during which the main part of this work was done.

\vspace{0.5cm}

\noindent {\bf Notations}: For a $\mathbb{Z}_2$-graded vector space $V$ over
the field $\mathbb{K}$ we write $V_{\overline{0}}$ for its even part and
$V_{\overline{1}}$ for its odd part, i.e. $V=V_{\overline{0}}\oplus
V_{\overline{1}}$. An element $X$ of $V$ is called homogeneous iff
$X\in V_{\overline{0}}$ or $X\in V_{\overline{1}}$. In this work, all
elements are supposed to be homogeneous unless otherwise stated.
For a homogeneous element $X$ we shall use the standard notation
$|X|\in\mathbb{Z}_2=\{\overline{0},\overline{1}\}$ to indicate its degree,
i.e. whether it is contained in the even part ($|X|=\overline{0}$) or in the
odd part ($|X|=\overline{1}$). Moreover, for two integers $k,l$ we denote
their images in $\mathbb{Z}_2$ by $\overline{k},\overline{l}$, and we use
the well-defined notation $(-1)^{\overline{k}\overline{l}}$ for
$(-1)^{kl}\in\{-1,1\}$.\\
All Lie superalgebras in this paper are finite-dimensional.

\section{Preliminaries}

\begin{defi}
{\rm  Let $\mathfrak{g}$ be a Lie superalgebra and let $B$
be a  bilinear form on $\mathfrak{g}$.
 \begin{enumerate}
\item[i)]
 $B$ is called  {\it supersymmetric }
 if  $B(X, Y) = (-1)^{\vert X\vert \vert Y\vert} B(Y,X),
 ~~\forall X~ \in \mathfrak{g}_{\vert X\vert},
  ~\forall~ Y \in \mathfrak{g}_{\vert Y\vert}$.
\item[ii)] $B$ is called    {\it invariant}  if $B([X, Y], Z) = B(X, [Y, Z]),
~~\forall ~ X, Y, Z  ~\in ~ \mathfrak{g}$.
 \item[iii)] $B$ is called   {\it even } if $B(X, Y) = 0, ~~ \forall~ X \in
\mathfrak{g}_{\bar 0},  ~\forall~ Y \in   \mathfrak{g}_{\bar 1}$.
\end{enumerate}}
\end{defi}

\begin{defi}
{\rm
\begin{enumerate}
\item[i)]  A Lie superalgebra   $\mathfrak{g}$ is called
{\it quadratic} if there exists a bilinear form $B$ on $\mathfrak{g}$ such that $B$ is
supersymmetric, even, non-degenerate and invariant. In this case, $B$ is
called an {\it invariant scalar product} on $\mathfrak{g}$.
 \item[ii)] Let $(\mathfrak{g},B)$ be a quadratic Lie superalgebra.
\begin{enumerate}
\item[1)] A graded ideal $\mathfrak{I}$ of $\mathfrak{g}$ is called  non-degenerate 
(resp. degenerate) if the restriction of $B$ to $\mathfrak{I}\times \mathfrak{I}$ is
a non-degenerate (resp. degenerate) bilinear form.

\item[2)] The quadratic Lie superalgebra $(\mathfrak{g},B)$ is called {\it  irreducible} if $\mathfrak{g}$ contains no
non-degenerate graded ideal other than $\{0\}$ and $\mathfrak{I}$.

 \item[3)] We say that a graded ideal $\mathfrak{I}$ of $\mathfrak{g}$
 is {\it irreducible} if
 $\mathfrak{I}$ is non-degenerate and $\mathfrak{I}$ contains no non-degenerate
 graded ideal of $\mathfrak{g}$ other than $\{0\}$ and $\mathfrak{I}$.
 \item[4)] A graded ideal $\mathfrak{I}$ of $\mathfrak{g}$ will be said to be {\it isotropic}
  if
 $B(\mathfrak{I},\mathfrak{I})=\{0\}$.
\end{enumerate}
\end{enumerate}}
\end{defi}

\begin{lema} \label{nacho}  Let $(\mathfrak{g},B)$ be a quadratic Lie superalgebra.
 Let $\mathfrak{I}$ be a graded ideal of $\mathfrak{g}$, we denote by
 $\mathfrak{I}^{\perp}$
 the orthogonal space of $\mathfrak{I}$ with respect to $B$.
\begin{enumerate}
\item[i)] $\mathfrak{I}^{\perp}$  is a graded ideal of $\mathfrak{g}$ and
  $[\mathfrak{I},\mathfrak{I}^{\perp}]= \{0\}$.
\item[ii)] $\mathfrak{I}$ is  non-degenerate if and only if $\mathfrak{I}^{\perp}$ is
 non-degenerate.
 \item[iii)] If $[\mathfrak{g},\mathfrak{I}]= \mathfrak{I}, $
 then $\mathfrak{I}^{\perp}= \mathcal{C}_\mathfrak{g}(\mathfrak{I})$,
 where $\mathcal{C}_\mathfrak{g}(\mathfrak{I})$ is the centralizer of
$\mathfrak{I}$  in $\mathfrak{g}$. Moreover,
$[\mathfrak{g},\mathfrak{g}]^{\perp}= \mathfrak{z}(\mathfrak{g})$.
 \item[iv)] If $\mathfrak{I}$ is  non-degenerate, then
 $(\mathfrak{I},{\tilde B}= B_{\vert_{\mathfrak{I}\times \mathfrak{I}}})$
 is a quadratic Lie
 superalgebra.  Moreover the quadratic  Lie superalgebra $(\mathfrak{I},{\tilde B})$ is
irreducible if and only if $\mathfrak{I}$ is an irreducible graded ideal of $\mathfrak{g}$.
 \item[v)] If $\mathfrak{H}$ is a semisimple graded ideal of $\mathfrak{g}$,
 then $\mathfrak{H}$ is
non-degenerate and $[\mathfrak{H},\mathfrak{H}]= \mathfrak{H}$.
\end{enumerate}
\end{lema}

The following Proposition reduces the study of quadratic Lie superalgebras to those
having no non-degenerate graded ideal other than $\{0\}$ and $\mathfrak{I}$.

\begin{prop}\label{saidpre} Let $(\mathfrak{g},B)$ be a quadratic Lie superalgebra. Then,
${\mathfrak{g}}= \oplus_{i=1}^{n}{\mathfrak{g}}_i$, where
\begin{enumerate}
\item[i)] ${\mathfrak{g}}_i$ is a non-degenerate graded ideal,
    for all $i\in \{1,\dots,n\};$

\item[ii)] $({\mathfrak{g}}_i,B_i= B_{\vert_{{\mathfrak{g}}_i\times {\mathfrak{g}}_i}})$
    is a quadratic Lie
    superalgebra,  for all $i\in \{1,\ldots,n\};$

\item[iii)] $B({\mathfrak{g}}_i,{\mathfrak{g}}_j)= \{0\}$,
  for all $i,j \in \{1,\ldots,n\}$.

\end{enumerate}
\end{prop}
A proof of both, Lemma \ref{nacho} and Proposition \ref{saidpre}, can be found for instance in \cite{Be18}.

\section{Notion of generalized double extension of quadratic Lie superalgebras}

>From now on, if ${\mathfrak g}$ is a Lie superalgebra, we will denote by $\mbox{\rm Der}({\mathfrak g})$ the Lie 
superalgebra of all its superderivations and by $\mbox{\rm End}({\mathfrak g})$ that of its linear endomorphisms. The symbol $\sum_{cyclic}$ will be frequently used to denote
cyclic sumation on a triple $X,Y,Z.$

\subsection{Generalized semi-direct product of Lie superalgebras}

Let $({\mathfrak{g}},[~,~]_{\mathfrak{g}})$ and $({\mathfrak{H}},[~,~]_{\mathfrak{H}})$
be two
Lie superalgebras, $F: {\mathfrak{g}}
\rightarrow \mbox{\rm Der}({\mathfrak{H}})$ be an even linear map (not necessarily a morphism of Lie
superalgebras), and $L: {\mathfrak{g}}\times {\mathfrak{g}}
\rightarrow {\mathfrak{H}}$ be an even superantisymmetric bilinear map such that
the following equations are satisfied:
\be \label{hom}
  [F(X),F(Y)]-F([X,Y]_{\mathfrak{g}})= \mbox{\rm ad}_{\mathfrak{H}}L(X,Y),
\ee
\be \label{cyc}
  \sum_{cyclic}(-1)^{|X||Z|} \biggl(F(X)\Bigl(L(X,Y)\Bigl) - L([X,Y]_{\mathfrak{g}},Z)
  \biggl)= 0,~ \forall
(X,Y,Z) \in  \mathfrak{g}_{\vert X\vert} \times
\mathfrak{g}_{\vert Y\vert}\times \mathfrak{g}_{\vert Z\vert}.
\ee
 We define the following product $[~,~]$ on  the
 $\mathbb{Z}_2-$graded $\mathbb{K}-$vector space
 $\mathfrak{G}= {\mathfrak{g}}\oplus {\mathfrak{H}}:$

\be
  [X+h,Y+l]=  [X,Y]_{\mathfrak{g}} + F(X)\Bigl(l\Bigl) -(-1)^{xy}F(Y)\Bigl(h\Bigl) + L(X,Y)
    + [h,l]_{\mathfrak{H}},
\ee
for all
$(X+h,Y+l) \in  \mathfrak{G}_{\vert X\vert} \times \mathfrak{G}_{\vert Y\vert}$.

Using (\ref{hom}) and (\ref{cyc}), we can see that the product $[~,~]$ defined a
Lie superalgebra structure on $\mathfrak{G}$.  The Lie superalgebra
$\mathfrak{G}$ will be called the generalized
semi-direct product of $\mathfrak{H}$ by $\mathfrak{g}$ by means of $(F,L)$ (See \cite{al}
for more informations on this type of extension).

\begin{rema} {\rm In particular, if $L= 0$ then $\mathfrak{G}$ is the   semi-direct product
of $\mathfrak{H}$ by $\mathfrak{g}$ by means of $F$ (cf. \cite{S}, Chapter III, §1).}
\end{rema}

\subsection{Generalized double extension of quadratic Lie superalgebras}

Let $({\mathfrak{g}}_1,[~,~]_1,B_1)$ a quadratic Lie superalgebra. We denote
by ${\rm Der}_a({\mathfrak{g}}_1)$ the Lie superalgebra of all
superderivations of ${\mathfrak{g}}_1$ which are superantisymmetric with respect
to $B_1$. Let ${\rm Out}_a({\mathfrak{g}}_1)$ be the quotient of
${\rm Der}_a({\mathfrak{g}}_1)$
by all inner derivations of ${\mathfrak{g}}_1$ which is known to be a Lie
superalgebra.
Let $({\mathfrak{g}}_2,[~,~]_2)$ be an arbitrary Lie superalgebra of finite dimension
and let $B_2$ be a supersymmetric invariant (not necessarily nondegenerate) bilinear
form on ${\mathfrak{g}}_2$.

\noindent Let
\be\label{phieq}
  \phi:{\mathfrak{g}}_2  \ra  {\rm Der}_a({\mathfrak{g}}_1)
\ee
be an even linear map, and
\be \label{psieq}
  \psi:{\mathfrak{g}}_2\times{\mathfrak{g}}_2  \ra  {\mathfrak{g}}_1
\ee
be an even bilinear superantisymmetric map which satisfy the following twisted
morphism condition
for all homogeneous elements $X,Y,Z\in {\mathfrak{g}}_2$:
\be\label{morpheq}
      [\phi(X),\phi(Y)]-\phi([X,Y]_2)= {\rm ad}_{{\mathfrak{g}}_1}(\psi(X,Y)).
\ee
Note that this condition implies that the composed map
$\tilde{\phi}:\mathfrak{g}_2\stackrel{\phi}{\rightarrow}
\mathrm{Der}_a(\mathfrak{g}_1)\rightarrow \mathrm{Out}_a(\mathfrak{g}_1)$
is a morphism of Lie superalgebras.
\begin{lema}
 With the above notation, the maps $\phi$ and $\psi$ satisfy the following equation:
 \be
  \sum_{cyclic}(-1)^{{\vert Z\vert}{\vert X\vert}}{\rm ad}_{{\mathfrak{g}}_2}\Big(
                        \phi(X)(\psi(Y,Z))- \psi([X,Y]_2,Z)\Big) = 0.
 \ee
\end{lema}
\dem
The Lemma easily follows by observing that the graded cyclic sum of
$[[\phi(X),\phi(Y)],\phi(Z)]$
vanishes by means of the super Jacobi identity
and by using twice equation (\ref{morpheq}).

This Lemma motivates the following, slightly stronger condition on $\phi$ and $\psi$:
\be \label{jaceq}
   \sum_{\rm cyclic} (-1)^{{\vert Z\vert}{\vert X\vert}}\big(
                        \phi(X)(\psi(Y,Z))- \psi([X,Y]_2,Z)\big)= 0.
\ee
Moreover, let
\be \label{chieq}
   \chi:{\mathfrak{g}}_2\times{\mathfrak{g}}_1  \ra  {\mathfrak{g}}_2^*
\ee
be the even bilinear map defined by the following equation for all homogeneous elements
$X\in{\mathfrak{g}}_1$ and $A,B\in {\mathfrak{g}}_2$:
\be
   \chi(A,X)(B):=-(-1)^{{\vert X\vert}{\vert B\vert}}B_1(\psi(A,B),X).
\ee

Denoting the \textit{coadjoint representation of ${\mathfrak{g}}_2$} on its dual by
$A.F:=-(1)^{{\vert A\vert}{\vert F\vert}}F\circ {\rm ad}_{{\mathfrak{g}}_2}(A)$ for all
homogeneous elements
$A\in{\mathfrak{g}}_2$ and $F\in {\mathfrak{g}}_2^*$, we finally need an even
bilinear superantisymmetric map
\be \label{weq}
   w:{\mathfrak{g}}_2\times {\mathfrak{g}}_2 \ra {\mathfrak{g}}_2^*
\ee
satisfying a \textit{twisted cocycle condition}
\be \label{wcon}
  \sum_{\rm cyclic} (-1)^{{\vert C\vert}{\vert A\vert}}\big( A.w(B,C) + w(A,[B,C]_2)
  + \chi(A,\psi(B,C)) \big)
                 =  0.
\ee
and a \textit{supercyclicity condition}
\be\label{wcyc}
    w(A,B)(C) = (-1)^{({\vert B\vert}+{\vert C\vert}){\vert A\vert}}w(B,C)(A),
\ee
for all homogeneous elements $A,B,C\in {\mathfrak{g}}_2$.
Let finally
\be\label{grossphieq}
   \Phi:{\mathfrak{g}}_1\times{\mathfrak{g}}_1  \ra  {\mathfrak{g}}_2^*
\ee
be the even superantisymmetric bilinear map defined by the following equation
for all homogeneous elements $X,Y\in{\mathfrak{g}}_1$ and $A\in {\mathfrak{g}}_2$:
\be \label{phicon}
  \Phi(X,Y)(A) := (-1)^{{\vert A\vert}({\vert X\vert}+{\vert Y\vert})}B_1(\phi(A)(X),Y).
\ee
We shall subsume these notions in a
\begin{defi}{\rm 
Let $({\mathfrak{g}}_1,B_1)$ be a quadratic Lie superalgebra, ${\mathfrak{g}}_2$ a
Lie superalgebra
and maps $\phi,\psi,w$ defined above satisfying the equations (\ref{morpheq}),
(\ref{jaceq}), (\ref{wcon}) and (\ref{wcyc}). We call
 $({\mathfrak{g}}_1,B_1,{\mathfrak{g}}_2,\phi,\psi,w)$
     a {\em context of generalized double extension} (of the quadratic Lie superalgebra
  $({\mathfrak{g}}_1,B_1)$ by the Lie superalgebra ${\mathfrak{g}}_2$.}
\end{defi}
We have the following
\begin{lema}\label{cocycle}
  Let $({\mathfrak{g}}_1,B_1,{\mathfrak{g}}_2,\phi,\psi,w)$ be a context of
  generalized double
  extension of the quadratic Lie superalgebra $({\mathfrak{g}}_1,B_1)$ by the
  Lie superalgebra
  ${\mathfrak{g}}_2$.
\\
 Then the maps $\Phi$ and $\chi$ satisfy the following
 equations  for all homogeneous elements
 $X,Y\in{\mathfrak{g}}_1$ and $A,B\in {\mathfrak{g}}_2$:
 \be \label{Phiphichi}
  \Phi(\phi(A)X,Y) + (-1)^{|A||X|}\Phi(X,\phi(A)Y)
    - A.\Phi(X,Y) -\chi(A,[X,Y]_1)
                                     =  0,
 \ee
and
 \bea \label{superchi}
  \chi([A,B]_2,X) - \chi(A,\phi(B)X) + (-1)^{|A||B|} \chi(B,\phi(A)X) & &
                                                        \nonumber \\
  ~~~~~-A.\chi(B,X) + (-1)^{|A||B|}B.\chi(A,X) +
             \Phi(\psi(A,B),X)  & = & 0 \label{chiphiPhi}
 \eea
 Moreover, $\Phi$ is an even  two-cocycle of the Lie superalgebra ${\mathfrak{g}}_1$ where
 ${{\mathfrak{g}}_2}^*$ is a trivial module for ${\mathfrak{g}}_1$, i.e. for three
 homogeneous
 elements $X,Y,Z\in {\mathfrak{g}}_1$ one has
 \be
    \sum_{\rm cyclic} (-1)^{|X||Z|}\Phi(X,[Y,Z])= 0.
 \ee
\end{lema}
\dem
 Using the definitions of the maps $\phi,\psi,\chi$, and $\Phi$
 the first two equations follow in a straight forward manner from
 the defining conditions (\ref{morpheq}) and (\ref{jaceq}), respectively.
 The cocycle condition is a consequence of the fact that $\phi({\mathfrak{g}}_2)$ is a
 subset of ${\rm Der}_a({\mathfrak{g}}_1)$   and has been shown in \cite{BB}.

We are now in the position to put the structure of a quadratic Lie superalgebra
on the vector space
${\mathfrak{g}}:={\mathfrak{g}}_2\oplus{\mathfrak{g}}_1\oplus{\mathfrak{g}}_2^*$:

\begin{theor} \label{ContextLie}
 Let $({\mathfrak{g}}_1,B_1,{\mathfrak{g}}_2,\phi,\psi,w)$ be a context of
  generalized double
  extension of the quadratic Lie superalgebra $({\mathfrak{g}}_1,B_1)$
  by the Lie superalgebra ${\mathfrak{g}}_2$.

 Let $A,B\in{\mathfrak{g}}_2$, $X,Y\in {\mathfrak{g}}_1$, and $F,H\in{\mathfrak{g}}_2^*$
 homogeneous
 elements such that $|A|=|X|=|F|$ and $|B|=|Y|=|H|$.
 We define the following bracket $[~,~]$ on the vector space ${\mathfrak{g}}$:
 \bea \label{crochet}
   [A+X+F,B+Y+H] & := & [A,B]_2  \nonumber \\
                 &    & +[X,Y]_1 + \Phi(X,Y) + \psi(A,B) + \phi(A)(Y)
                                    - (-1)^{|A||B|} \phi(B)(X) \nonumber \\
                 &    & + A.H - (-1)^{|A||B|}B.F
                          +\chi(A,Y)- (-1)^{|A||B|}\chi(B,X)\nonumber \\
                 &    &      ~~~       +w(A,B),
 \eea
 and the following bilinear form $B$ on ${\mathfrak{g}}$:
 \be \label{form}
   B(A+X+F,B+Y+H) := B_1(X,Y) +F(B) + (-1)^{|A||B|}H(A).
 \ee
 Then the triple $({\mathfrak{g}},[~,~],B)$ is a quadratic Lie superalgebra such that
 the subspace ${\mathfrak{g}}_2^*$ is an isotropic ideal of ${\mathfrak{g}}$ and the subspace ${\mathfrak{g}}_1\oplus {\mathfrak{g}}_2^*$ is the orthogonal space
         of ${\mathfrak{g}}_2^*$.

The quadratic Lie superalgebra ${\mathfrak{g}}$ is called the {\em generalized double extension}
of the quadratic Lie superalgebra $({\mathfrak{g}}_1,B_1)$ by the Lie superalgebra
${\mathfrak{g}}_2$ by means of $(\phi,\psi,w)$.
\end{theor}

\dem
By Lemma \ref{cocycle} we have
$\Phi \in (Z^2({\mathfrak{g}}_1,{\mathfrak{g}}_2^*))_{\bar 0}$, it follows
that we can
define the following Lie superalgebra structure on the $\mathbb{K}-$vector space
${\mathfrak{g}}_1\oplus{\mathfrak{g}}_2^*:$
\[
[X+F,Y+H]= [X,Y]_{{\mathfrak{g}}_1} + \Phi(X,Y),~
  \forall~ X+F\in {({\mathfrak{g}}_1\oplus{\mathfrak{g}}_2^*)}_{|X|},
 Y+H\in{({\mathfrak{g}}_1\oplus{\mathfrak{g}}_2^*)}_{|Y|}.
\]
This Lie superalgebra ${\mathfrak{g}}_1\oplus{\mathfrak{g}}_2^*$ is the central
extension of ${\mathfrak{g}}_1$ by
${\mathfrak{g}}_2^*$ by means of $\Phi$.\\
Now, if $A$ is an homogeneous
 element  of  ${\mathfrak{g}}_2$ we consider
 ${\tilde {\phi}}(A) \in
 \bigl({\rm End}({\mathfrak{g}}_1\oplus{\mathfrak{g}}_2^*)\bigl)_{|A|}$ defined by:

\be \label {deriv}
 {\tilde {\phi}}(A)\bigl(X+F\bigl):={\phi}(A)\bigl(X\bigl) +
\pi(A)\bigl(F\bigl) + \chi(A,X), ~\forall A+F \in {\mathfrak{g}}_1\oplus{\mathfrak{g}}_2^*,
\ee
where $\pi$ is the coadjoint representation of ${\mathfrak{g}}_2$. It is easy to see that
 equation
(\ref{Phiphichi}) and the fact that ${\phi}(A) \in {\rm Der}({\mathfrak{g}}_1)$ imply that
 ${\tilde{\phi}}(A)$ is a superderivation of Lie superalgebra
 ${\mathfrak{g}}_1\oplus{\mathfrak{g}}_2^*$.
 It follows that we have the even linear map
\[
 {\tilde {\phi}}:
   {\mathfrak{g}}_2 \ra {\rm Der}({\mathfrak{g}}_1\oplus{\mathfrak{g}}_2^*)
\]
defined by (\ref {deriv}) on the homogeneous elements of ${\mathfrak{g}}_2$.\\
Let us consider the even bilinear
superantisymmetric map
\[
  {\tilde {\psi}}:
  {\mathfrak{g}}_2\times{\mathfrak{g}}_2  \ra  {\mathfrak{g}}_1\oplus{\mathfrak{g}}_2^*
\]
defined by:
\[
  {\tilde {\psi}}(A,B):= {\psi}(A,B) + w(A,B), ~\forall
A \in ({\mathfrak{g}}_2)_{|A|}, B \in ({\mathfrak{g}}_2)_{|B|}.
\]
By the equations (\ref{morpheq}), (\ref {superchi}) we have

\be \label{psdg1}
  [{\tilde {\phi}}(A),{\tilde {\phi}}(B)]-{\tilde {\phi}}([A,B]_2)=
       {\rm ad}_{{\mathfrak{g}}_1 \oplus{\mathfrak{g}}_2^*}({\tilde {\psi}}(A,B)),
       ~\forall A \in
        {( {\mathfrak{g}}_2)}_{|A|}, B \in {({\mathfrak{g}}_2)}_{|B|}.
\ee
Moreover, equations (\ref{jaceq}), (\ref{wcon}) imply that

\be \label {psdg2}
\sum_{cyclic}(-1)^{|A||C|} \biggl({\tilde {\phi}}(A)\Bigl({\tilde {\psi}}(B,C)\Bigl) -
 {\tilde {\psi}}([A,B]_{\mathfrak{g}},C)  \biggl)= 0,~
\forall (A,B,C) \in
  {( {\mathfrak{g}}_2)}_{|A|}\times {( {\mathfrak{g}}_2)}_{|B|}\times
  {( {\mathfrak{g}}_2)}_{|C|}.
\ee
Consequently, equations (\ref {psdg1}), (\ref {psdg2}) imply that we can consider
the generalized semi-direct product
${\mathfrak{g}}:={\mathfrak{g}}_2\oplus{\mathfrak{g}}_1\oplus{\mathfrak{g}}_2^*$ of
${\mathfrak{g}}_1\oplus{\mathfrak{g}}_2^*$
by ${\mathfrak{g}}_2$ by means of $({\tilde {\phi}},{\tilde {\psi}})$.
It  is
easy to verify that the bracket of $\mathfrak{g}$ is exactly given by the
bracket $[~,~]$ of equation (\ref{crochet})
defined in
Theorem \ref{ContextLie}, and the  bilinear form $B$ as defined in (\ref {form}) is an
invariant scalar
product on $\mathfrak{g}$.  Finally, it is clear that the subspace
${\mathfrak{g}}_2^*$ is an isotropic
ideal of
${\mathfrak{g}}$, and that the  subspace ${\mathfrak{g}}_1\oplus {\mathfrak{g}}_2^*$
is the orthogonal space of
${\mathfrak{g}}_2^*$.

\begin{rema}
{\rm 
If ${\mathfrak{g}}$ is the generalized double extension of the
quadratic Lie superalgebra
$({\mathfrak{g}}_1,B_1)$ by the Lie superalgebra ${\mathfrak{g}}_2$ by means of
$(\phi,\psi,w)$ and if
 $B_2$ is  a supersymmetric invariant (not necessarily nondegenerate) bilinear form on
 ${\mathfrak{g}}_2$, then the following bilinear form $B'$ on ${\mathfrak{g}}$:
 \be
   B'(A+X+F,C+Y+H) := B_2(A,C)+B_1(X,Y)+F(C) + (-1)^{|A||C|}H(A),
\ee
is another invariant scalar product on ${\mathfrak{g}}$.}
\end{rema}


\subsection{Double extension and $T^*$-extension: Important particular cases}

Two particular cases of the generalized double extension should be mentioned: 
double extension and $T^*$-extensions. We first recall the notion of double extension of quadratic Lie
superalgebras as given in \cite{BB}:


\begin{theor} Let  $(\mathfrak{g}_1, B_1)$  be a quadratic Lie superalgebra,
$\mathfrak{g}_2$ a
Lie superalgebra and $\phi: \mathfrak{g}_2 \rightarrow
\mathrm{Der}_a(\mathfrak{g}_1)
~\subset~\mathrm{Der}(\mathfrak{g}_1)$  a morphism of Lie superalgebras.\\
Let  $\psi$  be the map from $\mathfrak{g}_1\times \mathfrak{g}_1$  to
$\mathfrak{g}_2^*$, defined by:
\[
 \psi(X, Y)(Z) = (-1)^{(|X|+|Y|)|Z|} B_1(\psi(Z)(X), Y)~~~~
   \forall X \in (\mathfrak{g}_1)_{|X|},
   ~\forall Y \in (\mathfrak{g}_1)_{|Y|},
   ~\forall Z \in (\mathfrak{g}_2)_{|Z|}.
\]
 Let   $\pi$  be the coadjoint representation  of  $\mathfrak{g}_2$.  Then  the
 $\mathbb{Z}_2-$graded vector space
 $\mathfrak{g} = \mathfrak{g}_2 \oplus \mathfrak{g}_1
 \oplus {\mathfrak{g}_2}^*$ with the product defined by
\begin{eqnarray*}
 [X_2 + X_1+ f, Y_2 + Y_1 + g] & = & [X_2, Y_2]_{\mathfrak{g}_2}
                                   + [X_1, Y_1]_{\mathfrak{g}_1}
                                   + \phi(X_2)(Y_1)
                                   - (-1)^{|X||Y|} \phi(Y_2)(X_1) \\
                               & &  +\pi(X_2)(g) -
  (-1)^{|X||Y|}\pi(Y_2)(f) + \psi(X_1, Y_1),
 \end{eqnarray*}
 (where  $|X|=|X_2|=|X_1|=|f|$ and $|Y|=|Y_2|=|Y_1|=|g|$)
 is a Lie superalgebra.

 Moreover, if  $\gamma$ is an invariant supersymmetric bilinear form on
 $\mathfrak{g}_2$,  then the bilinear form $T$ defined on  $\mathfrak{g}$ by
\[
   T(X_2 + X_1 + f, Y_2 + Y_1 + g) = B_1(X_1, Y_1) + \gamma(X_2, Y_2) +
                                       f(Y_2) + (-1)^{|X||Y|} g(X_2)
\]
is an invariant scalar product on $\mathfrak{g}$.

 The Lie superalgebra  $\mathfrak{g}$  is called a \textit{double extension of
 $({\mathfrak{g}}_1,B_1)$ by ${\mathfrak{g}}_2$ by means of $\phi$.}
\end{theor}

Let  $({\mathfrak{g}}_1,B_1,{\mathfrak{g}}_2,\phi,\psi,w)$ be a context of generalized
double extension of
the quadratic Lie superalgebra $({\mathfrak{g}}_1,B_1)$ by the Lie superalgebra
${\mathfrak{g}}_2$. If $\psi=0, w=0$ then
the generalized double extension ${\mathfrak{g}}$  of the quadratic Lie
superalgebra
$({\mathfrak{g}}_1,B_1)$ by the Lie superalgebra ${\mathfrak{g}}_2$ by means of
$(\phi,\psi,w)$ is actually the double extension of
$({\mathfrak{g}}_1,B_1)$ by ${\mathfrak{g}}_2$
by means $\phi$ in the sense of the Theorem.





\bigskip

Now, let $({\mathfrak{g}}_1,B_1,{\mathfrak{g}}_2,\phi,\psi=0,w)$ be a context of
generalized
double extension of the quadratic Lie superalgebra $({\mathfrak{g}}_1= 0,B_1= 0)$ by
the Lie superalgebra ${\mathfrak{g}}_2$.  Let ${\mathfrak{g}}$ be  the generalized
double extension
  of the quadratic Lie superalgebra $({\mathfrak{g}}_1= 0,B_1= 0)$ by the Lie superalgebra
  ${\mathfrak{g}}_2$ by means of $(\phi= 0,\psi= 0,w)$. The superalgebra $\mathfrak{g}$ will be
  called the
  \textit{$T^*$-extension of $\mathfrak{g}$ by means of $w$}, and we shall denote
  $\mathfrak{g}$ by the
  symbol $T^*_w {\mathfrak{g}}$ or $T^*{\mathfrak{g}}$. In this case,
  by Theorem \ref{ContextLie},
  we have the following Proposition.

\begin{prop} Let ${\mathfrak{g}}_2$ be a Lie
superalgebra,  ${\mathfrak{g}}_2^*$ its dual space, and $\pi$ its coadjoint representation.
Let $w$ be an even $2$-cocycle of ${\mathfrak{g}}_2$ with values in ${{\mathfrak{g}}_2}^*$.
Define the following
structures on the vector space ${\mathfrak{g}}:= {\mathfrak{g}}_2\oplus {\mathfrak{g}}_2^*$:
For a pair of homegeneous elements $~X+F, Y+H$ of $\mathfrak{g}$ of
degree $\vert X\vert=|F|$ (resp$.\vert Y\vert=|H|$), let
\[
  [X+F,Y+H]_{\mathfrak{g}}   := [X,Y]_{\mathfrak{g}}+w(X,Y)
   +\pi(X)(H)-(-1)^{\vert X \vert \vert Y \vert} \pi(Y)(F),
\]
and
\[
    B(X+F,Y+H) :=   F(Y)~+~(-1)^{|X||Y|}~H(X).
\]

Then $\mathfrak{g}$ endowed with the product $[\,,\,]_{\mathfrak{g}}$ is a Lie
superalgebra.\\

Moreover,  $({\mathfrak{g}},B)$ is a quadratic Lie superalgebra if and only if
$w$ is supercyclic, i.e.
\[
  w(X,Y)(Z)=~ (-1)^{\vert X\vert(\vert Y\vert+\vert Z\vert)}~w(Y,Z)(X),~~
  \forall~(X,Y,Z) \in {{\mathfrak{g}}_{\vert X\vert}\times {\mathfrak{g}}_{\vert Y\vert}
      \times{\mathfrak{g}}_{\vert Z\vert}}.
\]
We shall speak of the quadratic Lie superalgebra $({\mathfrak{g}},B)$
constructed out of $\mathfrak{g}$ and the supercyclic $2$-cocycle $w$ as a
{\em $T^*$-extension of $\mathfrak{g}$ by means $w$} and shall denote
${\mathfrak{g}}$ by the symbol
$T^*_w {\mathfrak{g}}$ or $T^* {\mathfrak{g}}$.
\end{prop}

\begin{rema} {\rm It is clear from the definition that if
$\mathfrak{g} =T^*_w (\mathfrak{g} _2)$,
then its dimension $n=\mathrm{dim}(\mathfrak{g})$ is even,
and $\mathfrak{I}=\mathfrak{g}_2^*$ is an
isotropic graded ideal of dimension $n/2$. Further, as a consequence
of Theorem \ref{conv} below, it is not difficult to see that the fact of being
even-dimensional and the existence of a graded ideal $\mathfrak{I}$ such that
$\mathfrak{I}=\mathfrak{I}^\perp$ characterize $T^*$-extensions.}
\end{rema}

It should be noticed that if $w$ is a supercyclic even 2-cocycle with values
in $\mathfrak{g} _2^*$ the the trilinear mapping
$\hat{w}: (\mathfrak{g}_2)^3\to \mathbb{K}$
given by $\hat{w}(X,Y,Z)=w(X,Y)Z$, turns out to be an even scalar 3-cocycle.
Actually, one gets in this way an isomorphism between the subspace of
$(Z^2(\mathfrak{g}_2,\mathfrak{g}_2^*))_{\bar{0}}$ composed of all supercyclic elements and
$(Z^3(\mathfrak{g}_2,\mathbb K))_{\bar{0}}$. Therefore one easily gets the following
result which gives a certain classification of the $T^*$-extensions of a given
Lie algebra in terms of the cohomology class of $\hat{w}$ (see \cite{BBBarXiv}).

\begin{prop} Let $\mathfrak{g}_2$ be a Lie superalgebra, $w_1,w_2$ two
    supercyclic even 2-cocycles with values in $\mathfrak{g}_2^*$, and consider an
    even
    superantisymmetric bilinear mapping
    $\varphi:\mathfrak{g}_2\times\mathfrak{g}_2\to\mathbb{K}$.\\
The map $S_\varphi:T^*_{w_1} (\mathfrak{g}_2)\to T^*_{w_2} (\mathfrak{g}_2)$ defined
by $S_\varphi (X+F)=X+\varphi(X,\cdot )+F$ defines an isometry of quadratic Lie
superalgebras if and only if $\hat{w}_1-\hat{w}_2$ is a coboundary.
\end{prop}

\section{ Inductive description of quadratic Lie  superalgebras}

In this section, we are going to give an inductive description of quadratic Lie
superalgebras  by using the notion of generalized  double extension of quadratic
Lie superalgebras.

\begin{theor} \label{conv}
Let  $({\mathfrak{g}},B)$  be a quadratic Lie superalgebra,
$\mathfrak{I}$ be an isotropic graded ideal of $\mathfrak{g}$, and
$\mathfrak{I}^{\perp}$ its orthogonal space. Define
\begin{enumerate}
 \item the quadratic Lie superalgebra
   $(\mathfrak{g}_1,B_1)$ with
   $\mathfrak{g}_1:=\mathfrak{I}^{\perp}/\mathfrak{I}$
   and $B_1$ the quadratic form induced by the restriction of $B$ to
   $\mathfrak{I}^{\perp}\times \mathfrak{I}^{\perp}$, and
 \item the Lie superalgebra $\mathfrak{g}_2:=\mathfrak{g}/\mathfrak{I}^{\perp}$.
\end{enumerate}
Then for any isotropic graded
vector subspace $\mathcal{V}$ of $\mathfrak{g}$ such that
${\mathfrak{g}}= \mathfrak{I}^{\perp}\oplus\mathcal{V}$ and any graded subspace
$\mathcal{A}$ of $\mathfrak{I}^{\perp}$ with
$\mathfrak{I}^\perp=\mathcal{A}\oplus \mathfrak{I}$ there is an even linear map
$\phi:\mathfrak{g}_2\to \mathrm{Der}_a(\mathfrak{g}_1)$, an even bilinear
superantisymmetric map
$\psi:\mathfrak{g}_2\times \mathfrak{g}_2\to \mathfrak{g}_1$, and an even
bilinear superantisymmetric map $w:\mathfrak{g}_2\times \mathfrak{g}_2\to
\mathfrak{g}_2^*$ such that
$({\mathfrak{g}},B)$ is isometric to the generalized double extension of the quadratic
Lie superalgebra
$(\mathfrak{g}_1,B_1)$ by $\mathfrak{g}_2$ by means
of $(\phi,\psi,w)$.
\end{theor}


\dem
It is clear that
$\mathfrak{I}\subseteq \mathfrak{I}^{\perp}$ since $\mathfrak{I}$ is isotropic.
Choose a graded vector subspace $\mathcal{A}$ of $\mathfrak{I}^\perp$ such
that $\mathfrak{I}^\perp=\mathcal{A}\oplus\mathfrak{I}$. Since
$\mathfrak{I}$ is the orthogonal space of $\mathfrak{I}^\perp$ it is
obvious that $\mathcal{A}$ is nondegenerate. Hence $\mathcal{A}^\perp$ is
nondegenerate, and we have $\mathcal{A}^\perp\cap \mathfrak{I}^\perp=\mathfrak{I}$.
Moreover, since $\mathcal{A}\cap\mathfrak{I}=\{0\}$ it follows
that $\mathfrak{g}=\mathcal{A}^\perp + \mathfrak{I}^\perp$
 Choose a  graded vector subspace
$\mathcal{V}$ of $\mathcal{A}^\perp\subseteq\mathfrak{g}$ such that
$\mathfrak{g}= \mathfrak{I}^{\perp}\oplus\mathcal{V}$. Hence $\mathcal{A}^\perp
=\mathcal{V}\oplus\mathfrak{I}$. Since the field $\mathbb{K}$ is algebraically closed
and of characteristic different from $2$ we may choose the graded subspace
$\mathcal{V}$ to be isotropic.

For any $A\in\mathfrak{g}=\mathcal{V}\oplus \mathcal{A}\oplus \mathfrak{I}$
denote by $A_{\mathcal{V}}$ its component in $\mathcal{V}$, by
$A_{\mathcal{A}}$ its component in $\mathcal{A}$, by
$A_{\mathfrak{I}}$ its component in $\mathfrak{I}$, and by
$A_{\mathfrak{I}^\perp}$ its component in $\mathfrak{I}^\perp$.
Clearly, $A=A_{\mathcal{V}}+A_{\mathcal{A}}+A_{\mathfrak{I}}
=A_{\mathcal{V}}+A_{\mathfrak{I}^\perp}$. Moreover, let
$\mathfrak{I}^\perp\rightarrow \mathfrak{g}_1:X\mapsto \overline{X}$ be
the natural projection which is a morphism of Lie superalgebras, and let
$K:\mathcal{V}\rightarrow \mathfrak{g}_2$ be the even linear map
$\mathcal{V}\hookrightarrow \mathfrak{g}\rightarrow \mathfrak{g}_2$ which
clearly is a linear isomorphism satisfying
\[
    K\big([V_1,V_2]_{\mathcal{V}}\big)=[KV_1,KV_2]_{\mathfrak{g}_1}
\]
for all $V_1,V_2\in\mathcal{V}$.
Using this notation we shall define the following three even maps
\beas
 \phi: \mathfrak{g}_2\rightarrow
 \mathrm{Hom}_{\mathbb{K}}(\mathfrak{g}_1,\mathfrak{g}_1) & : &
   KV\mapsto (\overline{X}\mapsto \overline{[V,X]}), \\
 \psi: \mathfrak{g}_2\times \mathfrak{g}_2 \rightarrow \mathfrak{g}_1
    & : & (KV_1,KV_2)\mapsto \overline{[V_1,V_2]_{\mathfrak{I}^\perp}}, \\
 w: \mathfrak{g}_2\times \mathfrak{g}_2 \rightarrow \mathfrak{g}_2^*
    & : & (KV_1,KV_2)\mapsto \big(KV_3\mapsto
    B([V_1,V_2],V_3)\big),
\eeas
which are well-defined since $K$ is a linear isomorphism and
$\mathfrak{I}^\perp$ is a graded ideal of $\mathfrak{g}$. \\
Since the ideals
$\mathfrak{I}^\perp$ and $\mathfrak{I}$ obviously are
$\mathfrak{g}$-modules by means of the adjoint representation,
it follows that their quotient ${\mathfrak g}_1$ is a
$\mathfrak{g}$-module on which $\mathfrak{g}$ acts by graded derivations.
Therefore the values of the linear map $\phi$ are graded derivations of
$\mathfrak{g}_1$, a fact which also follows directly by the graded Jacobi
identity of $\mathfrak{g}$. Moreover since the induced quadratic form
$B_1$ on $\mathfrak{g}_1$ is of the form $B_1(\overline{X},\overline{Y})
:=B(X,Y)$ for all $X,Y\in\mathfrak{I}^\perp$ we have for all
$V\in\mathcal{V}$
\beas
 B_1\big(\phi(KV)(\overline{X}),\overline{Y}\big)
   & = & B_1(\overline{[V,X]},\overline{Y})
   =B([V,X],Y)
   =-(-1)^{|V||X|}B(X,[V,Y])\\
   & = & -(-1)^{|V||X|}B_1\big(\overline{X},\phi(KV)(\overline{Y})\big),
\eeas
whence $\phi$ takes its values in $\mathrm{Der}_a(\mathfrak{g}_1)$ and satisfies
the defining condition (\ref{phieq}). \\
Moreover, $\psi$ is clearly an even
superantisymmetric bilinear map. The
fact that
$[V_1,V_2]=[V_1,V_2]_{\mathcal{V}}+[V_1,V_2]_{\mathfrak{I}^\perp}$
and the graded Jacobi identity on three elements $V_1,V_2,V_3\in
\mathcal{V}$ show that condition (\ref{jaceq}) is satisfied on $\phi$
and $\psi$.\\
Finally, it is obvious that $w$ is an even superantisymmetric bilinear
map. The supercyclic condition (\ref{wcyc}) for $w$ follows from the
invariance of $B$. Furthermore, for $V_1,V_2,V_3,V\in \mathcal{V}$ we
get for the map $\chi$ (see equation (\ref{chieq})):
\beas
  \chi\big(KV_1,\psi(KV_2,KV_3)\big)(KV)
   & = & - (-1)^{(|V_2|+|V_3|)|V|}B_1\big(\psi(KV_1,KV),\psi(KV_2,KV_3)\big)
                      \\
   & = & - (-1)^{(|V_2|+|V_3|)|V|}B([V_1,V]_{\mathfrak{I}^\perp},
                                [V_2,V_3]_{\mathfrak{I}^\perp})
\eeas
whence for one of the cyclic terms in condition (\ref{wcon}) we get
\beas
 \lefteqn{\Big((KV_1).\big(w(KV_2,KV_3)\big)\Big)(KV)
  +w(KV_1,[KV_2,KV_3])(KV)+\chi\big(KV_1,\psi(KV_2,KV_3)\big)(KV)  = }\\
  & &
  -(-1)^{(|V_2|+|V_3|)|V_1|}\Big(B([V_2,V_3],[V_1,V]_{\mathcal{V}})
                            +B([V_2,V_3]_{\mathcal{V}},[V_1,V])
  +B([V_2,V_3]_{\mathfrak{I}^\perp},[V_1,V]_{\mathfrak{I}^\perp})\Big)\\
  & & =-(-1)^{(|V_2|+|V_3|)|V_1|}\Big(B([V_2,V_3],[V_1,V])\Big)
\eeas
since $B([V_2,V_3]_{\mathcal{V}},[V_1,V])
=B([V_2,V_3]_{\mathcal{V}},[V_1,V]_{\mathfrak{I}^\perp})$ due to the fact
that $\mathcal{V}$ is isotropic. The graded cyclic sum of the preceding
terms vanishes, and thus condition (\ref{wcon}) holds thanks to the
invariance of $B$ and the graded Jacobi identity in $\mathfrak{g}$.

It follows that $\Big(\mathfrak{g}_1,B_1,\mathfrak{g}_2,\phi,\psi,w\Big)$
is a  context of generalized double extension of the quadratic Lie superalgebra
  $(\mathfrak{g}_1:={\mathfrak{I}^{\perp}{/}\mathfrak{I}},B_1)$ by the Lie superalgebra
  $\mathfrak{g}_2$ which is isomorphic to $\mathcal{V}$ as a graded
  vector space.
 Consider the generalized double extension
${\tilde \mathfrak{g}}= \mathfrak{g}_2\oplus\mathfrak{g}_1\oplus\mathfrak{g}_2^*$
of $(\mathfrak{g}_1,B_1)$ by $\mathfrak{g}_2$ by means of
$(\phi,\psi,w)$,
we denote by $\tilde B$ its  invariant scalar product  defined in Theorem
\ref{ContextLie}.\\
Let $\nabla:\mathfrak{I}\rightarrow \mathfrak{g}_2^*$ be the even linear
map
\[
    \nabla(X)(KV):=B(X,V).
\]
It is easy to verify in a long, but straightforward manner that the following linear map
\[
   \Pi: {\mathfrak{g}}= \mathcal{V}\oplus \mathcal{A}\oplus\mathfrak{I}   \rightarrow
{\tilde \mathfrak{g}}= \mathfrak{g}_2\oplus\mathfrak{g}_1\oplus\mathfrak{g}_2^*
\]
defined by
\[
   \Pi(V+A+X):= KV+\overline{A}+\nabla(X)
\]
for all $V\in\mathcal{V}$, $A\in\mathcal{A}$, and $X\in\mathfrak{I}$
is an isomorphism of Lie superalgebras. Moreover,
\[
  {\tilde B}(\Pi(X),\Pi(Y))= B(X,Y), ~~~\forall~ X,Y \in {\mathfrak{g}},
\]
 i.e. $\Pi$ is an isometry,
which proves the Theorem.
%
%


 \begin{coro}
 Let  $({\mathfrak{g}},B)$  be an irreducible quadratic
 Lie superalgebra which is neither
 simple nor the one-dimensional Lie algebra. Then, $({\mathfrak{g}},B)$ is a
 generalized double
 extension  of a quadratic Lie superalgebra $(\mathcal{A},T)$ by a simple Lie
superalgebra or by the one-dimensional Lie algebra or by the one-dimensional
odd Lie superalgebra  $\mathcal{N}=\mathcal{N}_{\bar 1}$.
\end{coro}

\dem  Since ${\mathfrak{g}}$  is neither a simple nor the one-dimensional Lie algebra,
then there exists a non-zero minimal graded ideal $\mathfrak{I}$ of ${\mathfrak{g}}$.
The fact that $({\mathfrak{g}},B)$ is an irreducible quadratic Lie superalgebra implies
that $\mathfrak{I}$ is isotropic. Therefore, by Theorem \ref{conv}, $({\mathfrak{g}},B)$
is
a generalized double extension of a quadratic Lie superalgebra
$({\mathfrak{I}^{\perp}{/}\mathfrak{I}},B_1)$ by the Lie superalgebra
$\mathfrak{g}/\mathfrak{I}^\perp$.
Since $\mathfrak{I}$ is a
minimal graded ideal of ${\mathfrak{g}}$, then $\mathfrak{I}^{\perp}$ is a maximal  graded
ideal of ${\mathfrak{g}}$. Consequently, ${\mathfrak{g}}{/}\mathfrak{I}^{\perp}$ is
either a simple
Lie superalgebra or the one-dimensional Lie algebra or the one-dimensional
odd Lie superalgebra  $\mathcal{N}=\mathcal{N}_{\bar 1}$.\\

Let $\mathcal{B}$ be the set consisting of $\{0\}$, the isometry classes of
quadratic simple Lie
superalgebras, and the one-dimensional Lie algebra.

\begin{theor}
Let $({\mathfrak{g}},B)$ be a quadratic Lie superalgebra.  Then, either
$\mathfrak{g}$ is an element of $\mathcal{B}$ or $({\mathfrak{g}},B)$ is obtained
by a sequence
of generalized double extensions by a simple Lie superalgebra or by the
one-dimensional Lie algebra or by the one-dimensional Lie superalgebra
$\mathcal{N}=\mathcal{N}_{\bar 1}$ and/or orthogonal direct sums of quadratic
Lie superalgebras from a finite number of elements of $\mathcal{B}$.
\end{theor}

\begin{rema}\rm
It can be easily seen that the Cartan superalgebra $W(n)= \mbox{Der} (\bigwedge V)$, where $V$ is a $n$-dimensional vector space, is not quadratic for $n\geq 3$ since for every even bilinear supersymmetric and invariant form $B$ on $W(n)$ the non-null subspace $W_{n-1}=\{D\in W(n)\,|\, D(V)\subset \bigwedge^{n}V\} $ is orthogonal with respect to $B$ to the whole $W(n)$.

In Proprosition 3 of \cite[page 187]{S} it is proved that if $V$ is a $n$-dimensional vector space with $n\geq 4$ then the gradation $S(V)=\oplus_{r=-1}^{n-2}S_r(V)$ verifies $[S_r(V),S_1(V)]= S_{r+1}(V)$ for all  $r\geq -1$. Therefore if $B$ is an even invariant bilinear supersymmetric form then
$B(S_{n-2}(V), S(V))= \{0\},$ which proves that $B$ is degenerate and hence $S(V)$ cannot be quadratic.

The same reasoning shows that $H(n)$ is not quadratic for $n\geq 5$
 since from Proprosition 7 in \cite[page 196]{S} one has $[H_r(V),H_1(V)]= H_{r+1}(V)$ for all $r\geq -1$ and hence    $ H_{n-3}$  must be orthogonal to $H(n)$ with respect to every even invariant bilinear supersymmetric form.                                                                                  

We conclude that, with the possible exception of the superalebras ${\tilde S}(2r)$, $r\geq 1$, a non-classical simple Lie superalgebra cannot be quadratic.
\end{rema}



\section{Solvable quadratic Lie superalgebras are generalized double extensions
by one-dimensional Lie superalgebras}

It has been proved in \cite{BB} that every $n$-dimensional irreducible quadratic
Lie superalgebra $\mathfrak{g}$ such that
$\mathfrak{z}(\mathfrak{g})\cap\mathfrak{g}_{\bar 0}\neq\{ 0\}$ is a
double extension of a $(n-2)$-dimensional quadratic Lie superalgebra by a
one-dimensional ideal. However, the condition
$\mathfrak{z}(\mathfrak{g})\cap\mathfrak{g}_{\bar 0}\neq\{ 0\}$
is not always satisfied (even in the nilpotent case) as  the following examples show:

\medskip

\noindent {\bf Examples} \begin{enumerate}
\item For each positive integer $n$, let  $\mathcal{T}(n)$  denote the linear space
of square matrices of order $n$ which are upper triangular and $\mathcal{N}(n)$ its
subspace of strict upper triangular matrices. It is easy to prove that
\[
  \mathfrak{g}(n)=\left.\left\{ \left(
               \begin{array}{cc}
                    A& B\\
                    C&D
               \end{array}\right)\, \right|\, A,C,D\in \mathcal{N}(n),
               B\in\mathcal{T}(n)\right\}
\]
is a Lie sub-superalgebra of $\mathfrak{gl}(n,n)$ and that
$\mbox{dim}\big(\mathfrak{z}\big(\mathfrak{g}(n)\big)\big)=1$,
$\mathfrak{z}\big(\mathfrak{g}(n)\big)\subset \mathfrak{g}(n)_{\bar 1}$, and
$~[\mathfrak{g}(n)_{\bar 1},\mathfrak{g}(n)_{\bar 1}]= \mathfrak{g}(n)_{\bar 0}$.
Let $\mathcal{E}_n=T^*_0(\mathfrak{g}(n))$ be the $T^*-$extension
of $\mathfrak{g}(n)$ by $w= 0$. It is clear that
$\mathcal{E}_n$ is nilpotent since so is $\mathfrak{g}(n)$. Its centre is given by
\[
   \mathfrak{z}(\mathcal{E}_n)
     =\mathfrak{z}\big(\mathfrak{g}(n)\big)\oplus \{f \in \mathfrak{g}(n)^*\, |\,
         f([\mathfrak{g}(n),\mathfrak{g}(n)])= \{0\}\}
\]
and since $[\mathfrak{g}(n)_{\bar 1},\mathfrak{g}(n)_{\bar 1}]
= \mathfrak{g}(n)_{\bar 0}$ one easily gets
that $\{0\}\neq\mathfrak{z}(\mathcal{E}_n)
\subset \mathfrak{g}(n)_{\bar 1}\oplus \mathfrak{g}(n)^*_{\bar 1}=
(\mathcal{E}_n)_{\bar 1}$.
\item The superalgebras constructed in the example above have dimension $2n(2n-1)$
and hence, the smallest one is the $12$-dimensional superalgebra $\mathcal{E}_2$.
M. Duflo drew our attention to the following $7$-dimensional example:
Consider the quadratic superspace $V$ where the even part is one-dimensional
and the odd part has dimension $2$ and let $\mathcal{N}$ be the standard nilpotent
sub-superalgebra of $\mathfrak{osp}(1,2)$. The superalgebra double extension of $V$ by
$\mathcal{N}$ has also its centre contained in the odd part.
\end{enumerate}

Obviously, the results in \cite{BB} cannot be applied in these examples. Further,
one can see that the second one gives a quadratic Lie superalgebra which cannot be
constructed as a (classical) double extension by a one-dimensional algebra. Thus,
the inductive classification of solvable quadratic Lie superalgebras requires the
use of generalized double extensions. We will prove in this section that, actually,
every such superalgebra may be obtained by sucessive generalized double extensions
by one-dimensional superalgebras.

\medskip

 It is easy to verify that
 $({\mathfrak{g}}_1,B_1,{\mathfrak{g}}_2:=\mathcal{N} ,\phi,\psi,w)$ is a context
 of generalized double
  extension of a quadratic Lie superalgebra $({\mathfrak{g}}_1,B_1)$  by the
  one-dimensional
  Lie superalgebra $\mathcal{N}=\mathcal{N}_{\bar 1}= \mathbb{K}e$,
  if and only if
  $w=0$ and there exists
  $(D,X_0)\in
  [\mbox{\rm Der}_a({\mathfrak{g}}_1,B_1)]_{\bar 1}\times ({\mathfrak{g}}_1)_{\bar 0}$
  such that:
\[
  \phi(e)= D,\,\,\psi(e,e)= X_0,\,\, D(X_0)= 0,\,\, B_1(X_0,X_0)= 0\,\,
\mbox{and}\,\, D^2= {1\over 2}\mbox{ad}_{{\mathfrak{g}}_1}X_0.
\]

If ${\mathfrak{g}}$ is  the generalized double extension   of the quadratic Lie
superalgebra $({\mathfrak{g}}_1,B_1)$ by the Lie superalgebra
${\mathfrak{g}}_2:=\mathcal{N}$ by means
of $(\phi,\psi,w:= 0)$, then the bracket $[\,,\,]$ on
${\mathfrak{g}}= \mathbb{K}e\oplus {\mathfrak{g}}_1 \oplus \mathbb{K}e^*$
(where $\{e^*\}$ is the dual basis of $\{e\}$) is defined by:
\[
  ~[e,e]= X_0,\, [e^*,{\mathfrak{g}}]= \{0\},
\]
\[
    ~[e,X]= D(X) - B_1(X,X_0) e^*,
\]
\[
     ~[X,Y]= [X,Y]_{{\mathfrak{g}}_1} - B(D(X),Y) e^*,\,
      \forall X \in {\mathfrak{g}}_{\vert X\vert},\, Y\in {\mathfrak{g}},
\]
and the invariant scalar product $B$ on $\mathfrak g$ is defined by:
\[
  B(e^*,e)= 1,\, B_{{\mathfrak{g}}_1\times{\mathfrak{g}}_1}:= B_1,
\]
\[
 B(x,e)= B(x,e^*)= B(e^*,e^*)= B(e,e)= 0,\,\forall X \in {\mathfrak{g}}_1.
\]

\begin{prop}\label{CI}
Let  $({\mathfrak{g}},B)$  be an irreducible  quadratic Lie superalgebra.
If $\mathfrak{z}({\mathfrak{g}})\cap {\mathfrak{g}}_{\bar 1}\not= \{0\}$,
 then $({\mathfrak{g}},B)$ is a generalized
double extension of a quadratic Lie superalgebra $(\mathcal{A},T)$ by the
one-dimensional Lie superalgebra  $\mathcal{N}=\mathcal{N}_{\bar 1}$.
\end{prop}
\dem  Suppose that $\mathfrak{z}({\mathfrak{g}})\cap {\mathfrak{g}}_{\bar 1}\not= \{0\}$.
Let $X \in \mathfrak{z}({\mathfrak{g}})\cap {\mathfrak{g}}_{\bar 1}\x\{0\}$.
Then $\mathfrak{I}= \mathbb{K}X$
is an isotropic graded ideal of $\mathfrak{g}$. Therefore there exists
$Y \in {\mathfrak{g}}_{\bar 1}$
such that $B(X,Y)= 1$, and $B(Y,Y)=0$. It follows that
$\mathfrak{g}= \mathfrak{I}^{\perp}\oplus \mathbb{K}Y$.
Then, by Theorem \ref{conv},
$({\mathfrak{g}},B)$ is a generalized double extension of a quadratic Lie superalgebra
$(\mathcal{A},T)$ by the one-dimensional Lie superalgebra
$\mathcal{N}=\mathcal{N}_{\bar 1}$.

\begin{coro} Let  $({\mathfrak{g}},B)$  be an irreducible quadratic Lie superalgebra such
that $\mathfrak{g}$ is not the one-dimensional Lie algebra.
If $\mathfrak{g}$ is solvable, then
either $({\mathfrak{g}},B)$ is  a double extension   of a solvable quadratic Lie
superalgebra $(\mathcal{A},T)$ by   the one-dimensional Lie algebra or $({\mathfrak{g}},B)$
 is a generalized double extension  of a solvable quadratic Lie superalgebra
 $(\mathcal{H},U)$ by   the one-dimensional Lie superalgebra.
\end{coro}
\dem Since ${\mathfrak{g}}$ is solvable, then $\mathfrak{z}({\mathfrak{g}})\not= \{0\}$.
If $\mathfrak{z}({\mathfrak{g}})\cap {\mathfrak{g}}_{\bar 1}\not= \{0\}$,
(resp. $\mathfrak{z}({\mathfrak{g}})\cap {\mathfrak{g}}_{\bar 0}\not= \{0\})$, then,
by Corollary \ref{CI} (resp. by \cite{BB}, Corollary 1 of Proposition 3.2.3),
$({\mathfrak{g}},B)$ is a generalized double extension  of a  quadratic Lie
superalgebra $(\mathcal{H},U)$ by   the one-dimensional Lie superalgebra
(resp. $({\mathfrak{g}},B)$ is  a double extension   of a  quadratic Lie superalgebra
$(\mathcal{A},T)$ by   the one-dimensional Lie algebra) where
$\mathcal{H}= \mathfrak{I}^{\perp}/\mathfrak{I}$
(resp. $\mathcal{A}= \mathfrak{I}^{\perp}/\mathfrak{I}$) with $\mathfrak{I}= \mathbb{K}X$
where $X \in \mathfrak{z}({\mathfrak{g}})\cap {\mathfrak{g}}_{\bar 1}\x\{0\}$
(resp. $X \in \mathfrak{z}({\mathfrak{g}})\cap {\mathfrak{g}}_{\bar 0}\x\{0\}$).
The fact that
$\mathfrak{g}$ is solvable implies that its graded ideal is also solvable, it follows
that the Lie superalgebra $\mathcal{H}$ (resp. $\mathcal{A}$) is solvable.

\begin{coro}
Let  $({\mathfrak{g}},B)$ be a solvable quadratic Lie superalgebra but
neither $\{0\}$ nor the one-dimensional Lie algebra.  Then, $({\mathfrak{g}},B)$ is
obtained by a sequence of double extensions of solvable quadratic Lie superalgebras
by the one-dimensional Lie algebra and/or generalized double extensions of
solvable quadratic Lie superalgebras by the one-dimensional Lie superalgebra
and/or orthogonal direct sums of  solvable quadratic Lie superalgebras constructed in that way.
\end{coro}

\section{Every  Solvable Quadratic Lie Superalgebra is described by a $T^*-$extension}

The following central Theorem was communicated to us by
M.~Duflo. We have made its proof more elementary upon
circumventing the use of universal envelopping algebra which was used in
Duflo's original proof.

\begin{theor}{\bf (Duflo)}\label{DL}
Let $\mathfrak{g}$ be a solvable Lie superalgebra and  $\mathcal{V}$  be a
finite-dimensional
${\mathfrak{g}}$-module. If $\mathcal{V}$ is simple, non-zero and
isomorphic to its dual $\mathcal{V}^*$,
then $\mathcal{V}$ is the one-dimensional trivial $\mathfrak{g}$-module.
\end{theor}

\dem Let $\mathcal{V}$ be a simple module of $\mathfrak{g}$. We shall denote by $\rho$
  the module morphism $\mathfrak{g}\rightarrow
 \mathrm{Hom}_{\mathbb{K}}(V,V)$, and  by $\rho^*$ the contragredient module map.

\noindent $\mathbf{1.}$ We shall first deal with the case that
  $\mbox{dim}\mathcal{V}= 1$. It follows that $\mathcal{V}= \mathbb{K}e$. Let $\{f\}$
 be the dual basis vector of $\{e\}$. Then there is a $1$-form
 $\lambda\in\mathfrak{g}^*$ such that for each  homogeneous element $X$ of
 $\mathfrak{g}$ we have
 $\rho(X)e= \lambda(X)e$ and $\rho^*(X)f= -(-1)^{\vert X\vert\vert e\vert}\lambda(X)f$.
 Let $L: \mathcal{V}^* \rightarrow \mathcal{V}$ be an
 isomorphism of ${\mathfrak{g}}$-modules. Then for each homogeneous element $X$
 of $\mathfrak{g}$ the fact that $L\big(\rho^*(X)f\big)= \rho(X)\big(L(f)\big)$
 implies that $(1+(-1)^{\vert X\vert\vert e\vert})\lambda(X)= 0$. Consequently,
 $\lambda(X)= 0$
 if $\vert X\vert= 0$.  Since  $\rho(X)e= \lambda(X)e$, it follows that $\lambda(X)= 0$ if
 $\vert X\vert= 1$ because the linear map $\rho(X)$ changes parity and $e$ is homogeneous.
 We conclude that $\mathcal{V}$ is a trivial
 $\mathfrak{g}$-module.

\noindent We are going to prove this Theorem now by induction on the dimension of
             $\mathfrak{g}$. We can and shall suppose henceforth
             that $\dim \mathcal{V}\geq 2$.

\noindent $\mathbf{2.1}$ If $\mbox{dim}{\mathfrak{g}}= 0$, then
$\mathcal{V}$ is a simple $\mathbb{K}$-vector space, whence
$\dim\mathcal{V}=1$ which contradicts the above hypothesis.

\noindent $\mathbf{2.2}$ If $\mbox{dim}{\mathfrak{g}}= 1$, then
${\mathfrak{g}}= \mathbb{K}X$ where $X$ is a homogeneous element of $\mathfrak{g}$.
Since $\mathbb{K}$
is algebraically closed, it follows that $\rho(X)$ has a nonzero eigenvector
$e \in \mathcal{V}\x \{0\}$
such that  $\rho(X)e= \lambda e$ and $\lambda \in \mathbb{K}$.
Let $e=e_{\overline{0}}+e_{\overline{1}}$ with
$e_{\overline{0}}\in\mathcal{V}_{\overline{0}}$ and
$e_{\overline{1}}\in\mathcal{V}_{\overline{1}}$. In case $X$ is even it
follows that both $e_{\overline{0}}$ and $e_{\overline{1}}$ are
eigenvectors of $X$ for the eigenvalue $\lambda$. Since at least one of
them is nonzero, there would exist a proper submodule of $\mathcal{V}$
of dimension $1$ which contradicts $\dim \mathcal{V}\geq 2$. In case
$X$ is odd, we have $\rho(X)(e_{\overline{0}})=\lambda e_{\overline{1}}$ and
$\rho(X)(e_{\overline{1}})=\lambda e_{\overline{0}}$. Therefore the subspace
generated by $e_{\overline{0}}$ and $e_{\overline{1}}$
is a submodule of $\mathcal{V}$, hence equal to it by simplicity of
$\mathcal{V}$. If $e_{\overline{0}}$ or $e_{\overline{1}}$ was zero, there would again be a
onedimensional submodule, contradiction. It follows
that $e_{\bar 0}\not= 0$ and $e_{\bar 1}\not= 0$, whence
$V$ is $2$-dimensional and spanned by $e_{\overline{0}}$ and $e_{\overline{1}}$.
Consider the dual basis $\{f_{\bar 0} ,f_{\bar 1}\}$ of $\mathcal{V}^*$
associated to
$\{e_{\bar 0},e_{\bar 1}\}$.  It follows that
$\rho^*(X)f_{\bar 0}=-\lambda f_{\bar 1}$ and $\rho^*(X)f_{\bar 1}= \lambda f_{\bar 0}$.
Since $\mathcal{V}$ and its dual $\mathcal{V}^*$ are isomorphic
${\mathfrak{g}}$-modules, we
can consider an isomorphism of ${\mathfrak{g}}$-modules
$L:\mathcal{V}^*\rightarrow  \mathcal{V}$. Define matrix elements
$L_{\overline{a}~\overline{b}}\in\mathbb{K}$ of $L$ by
$L(f_{\overline{b}}=L_{\overline{0}~\overline{b}}e_{\overline{0}}
+L_{\overline{1}~\overline{b}}e_{\overline{1}}$ for all
$\overline{a},~\overline{b}~\in \mathbb{Z}_2$. A little computation shows that
the isomorphism condition
$L\circ \rho^*(X)=\rho(X)\circ L$ implies that $\lambda=0$. But then for instance
$\mathbb{K}e_{\overline{0}}$ would be one-dimensional submodule, contradiction.

\noindent $\mathbf{2.3}$
Suppose now that $\mbox{dim}{\mathfrak{g}}\geq 2$, and the Theorem is true for all
solvable Lie
superalgebras $\mathcal{H}$ such that  $\mbox{dim}\mathcal{H}< \mbox{dim}{\mathfrak{g}}$.
 The fact that $\mathfrak{g}$ is solvable
 implies
 that there exists a graded ideal $\mathfrak{h}$ of $\mathfrak{g}$ of codimension $1$
 such that $[\mathfrak{g},\mathfrak{g}]\subset \mathfrak{h}\subset
 \mathfrak{g}$. There also exists a homogeneous element $T\in\mathfrak{g}$
 with $\mathfrak{g}=\mathfrak{h}\oplus \mathbb{K}T$ (direct sum of vector
 spaces). Moreover, let $\mathcal{W}\subset \mathcal{V}$ be a simple
 $\mathfrak{h}$-submodule of $\mathcal{V}$. Clearly, $\mathcal{W}$ is not
 equal to $\mathcal{V}$ since otherwise $\mathcal{W}=\mathcal{V}$ would be
 one-dimensional thanks to the induction hypothesis applied to
 $\mathfrak{h}$: this would contradict
 the  assumption $\mbox{dim}\mathcal{V}\geq 2$. Finally, for any nonnegative integer
 $i$ let $\mathcal{W}_i$ denote the following vector subspace of $\mathcal{V}$:
 \begin{equation}\label{EqDefWi}
   \mathcal{W}_i:=\mathcal{W}+\rho(T)\mathcal{W}+\cdots+\rho(T)^i\mathcal{W}.
 \end{equation}
 For strictly negative $i$ we set $\mathcal{W}_i:=\{0\}$. We are going to
 show the following statements:
 \begin{eqnarray}
   & & \mbox{There is a strictly positive integer}~M
             ~\mbox{such that}~\rho(T)\mathcal{W}_M\subset \mathcal{W}_M
             \nonumber \\
   & &             ~~~\mbox{and}~
                 \rho(T)\mathcal{W}_{M-1}\not\subset\mathcal{W}_{M-1};
          \mbox{moreover if}~|T|=\overline{1}~\mbox{then}~M=1.
                       \label{EqAussage1} \\
   & & \mathcal{W}_i~\mbox{is a}~\mathfrak{h}\mbox{-module for all}~
                           i\in\mathbb{N},
                \mbox{and}~\mathcal{W}_M=\mathcal{V}.
                      \label{EqAussage2} \\
   & & \mbox{The}~\mathfrak{h}\mbox{-module}~\mathcal{W}_i/\mathcal{W}_{i-1}~
      \mbox{is isomorphic to}~\mathcal{W}~\mbox{for
                           all integers}~0\leq i\leq M
                           \label{EqAussage3}
 \end{eqnarray}
 Indeed, to show statement (\ref{EqAussage1}), it is clear that
 $\rho(T)\mathcal{W}_i\subset \mathcal{W}_{i+1}$,
 and since $\mathcal{V}$ is finite-dimensional there is a nonnegative integer
 $N$ such that $\mathbf{1}_\mathcal{V},\rho(T),\ldots,\rho(T)^{N+1}$ are
 linearly dependent which implies that
 $\rho(T)\mathcal{W}_N\subset\mathcal{W}_N$. Then $M$ will be the minimum
 of all these integers. The case $M=0$ would imply that
 $\mathcal{W}_0=\mathcal{W}$ is a $\mathfrak{g}$-module contradicting the
 simplicity of the $\mathfrak{g}$-module $\mathcal{V}$ whence $M\geq 1$.
 Note that for odd $T$ we have $\rho([T,T])=2\rho(T)^2$ with
 $[T,T]\in [\mathfrak{g},\mathfrak{g}]\subset\mathfrak{h}$ showing $M=1$
 since $\mathcal{W}$ is a $\mathfrak{h}$-module.\\
 Statement (\ref{EqAussage2}) follows for odd $T$ immediately from the
 representation identity (for all homogeneous $x\in\mathfrak{g}$)
 \begin{equation}\label{EqRepIdentity}
   \rho(x)\rho(T)= \rho([x,T]) + (-1)^{|T||x|}\rho(T)\rho(x)
 \end{equation}
 since $M=1$ in that case, and $W_i=W_M$ for all integers $i\geq M$.
 For even $T$ we use the following identity
 \begin{equation}\label{EqRepIdentityVersioni}
   \rho(x)\rho(T)^i= \sum_{k=0}^i{i \choose k}
                       \rho(T)^k
   \rho(\underbrace{[\ldots[x,T]\ldots,T]}_{i-k\mathrm{~brackets}})
   ~~\forall~x\in\mathfrak{g}~\mbox{and}~\forall~i\in\mathbb{N},
 \end{equation}
 (deduced from the representation identity (\ref{EqRepIdentity}) by
 induction) for the particular case $x\in\mathfrak{h}$ to prove the
 statement since the iterated brackets $[\ldots[x,T]\ldots,T]$ are
 then contained in the ideal $\mathfrak{h}$. Since the $\mathfrak{h}$-module
 $\mathcal{W}_M$ is also stable by $\rho(T)$ according to (\ref{EqAussage1})
 it is a nonzero $\mathfrak{g}$-submodule of the simple $\mathfrak{g}$-module
 $\mathcal{V}$ and therefore equal to $\mathcal{V}$.\\
 For statement (\ref{EqAussage3}) consider the linear map
 \[
    \Phi_i:\mathcal{W}\rightarrow \mathcal{W}_i/\mathcal{W}_{i-1}:
             w\mapsto (-1)^{|T||w|}\rho(T)^i(w)+\mathcal{W}_{i-1}.
 \]
 for homogeneous $w\in \mathcal{W}$ and all integers $0\leq i\leq M$.
 Since $\mathcal{W}_i=\rho(T)^i\mathcal{W}+\mathcal{W}_{i-1}$ it follows
 that $\Phi_i$ is surjective. For odd $T$ it suffices to use the
 representation identity (\ref{EqRepIdentity}) to show that $\Phi_i$ is a
 morphism of $\mathfrak{h}$-modules in view of the fact that $M=1$,
 whereas for even $T$ we use identity (\ref{EqRepIdentityVersioni})
 to prove that each $\Phi_i$ is a morphism of $\mathfrak{h}$-modules.
 In both cases the kernel of $\Phi_i$ is an $\mathfrak{h}$-submodule of
 the simple $\mathfrak{h}$-module $\mathcal{W}$ so it is either  equal to
 $\mathcal{W}$ or equal to $\{0\}$: in the first case $\Phi_i$ would be
 the zero map implying $\mathcal{W}_i=\mathcal{W}_{i-1}$ which is a
 contradiction to the definition of the integer $M$. It follows that
 $\Phi_i$ is injective and hence an isomorphism of $\mathfrak{h}$-modules.

 We continue the proof of the Theorem: statement (\ref{EqAussage3})
 implies in particular for $i=M$ that the $\mathfrak{h}$-module $\mathcal{W}$ is
 isomorphic to the quotient module $\mathcal{V}/\mathcal{W}_{M-1}$. Hence
 the dual $\mathfrak{h}$-module $\mathcal{W}^*$ is isomorphic to the
 $\mathfrak{h}$-module $(\mathcal{V}/\mathcal{W}_{M-1})^*$ which in turn
 is canonically isomorphic the the $\mathfrak{h}$-submodule
 \[
     \mathcal{W}_{M-1}^{\mathrm{ann}}:=\{f\in V^*~|~f(v)=0~\forall~v\in
                                   \mathcal{W}_{M-1}\}\subset\mathcal{V}^*
 \]
 by means of the map
 $\psi:\mathcal{W}_{M-1}^{\mathrm{ann}}\to (\mathcal{V}/\mathcal{W}_{M-1})^*$
 defined by $\psi(f)\big(v+\mathcal{W}_{M-1}\big):=f(v)$ (for all
 $f\in \mathcal{W}_{M-1}^{\mathrm{ann}}$ and $v\in\mathcal{V}$). Consider
 now an isomorphism of $\mathfrak{g}$-modules $L:\mathcal{V}^*\to
 \mathcal{V}$, and let $\tilde{\mathcal{W}}$ be the
 $\mathfrak{h}$-submodule of $\mathcal{V}$ defined by
 $\tilde{\mathcal{W}}:=L\big(\mathcal{W}_{M-1}^{\mathrm{ann}}\big)$. Since
 for each nonnegative integer $i$ we have $\mathcal{W}_i\subset
 \mathcal{W}_{i+1}$ it follows that there is a smallest integer $k$ such
 that $\tilde{\mathcal{W}}\subset \mathcal{W}_k$ and
 $\tilde{\mathcal{W}}\not\subset \mathcal{W}_{k-1}$. The image of
 $\tilde{\mathcal{W}}$ under the $\mathfrak{h}$-module map $L':\mathcal{W}_{k}\to
 \mathcal{W}_k/\mathcal{W}_{k-1}\cong W$ is an $\mathfrak{h}$-submodule of
 $\mathcal{W}$, hence it is either equal to $\{0\}$ or to $\mathcal{W}$.
 If the image was equal to $\{0\}$, then $\tilde{\mathcal{W}}\subset
 \mathcal{W}_{k-1}$, a contradiction, and hence the image is equal to
 $\mathcal{W}$. Since
 \[
    \dim \tilde{\mathcal{W}}=\dim(\mathcal{W}_{M-1}^{\mathrm{ann}})
      =\dim W^* =\dim W
 \]
 it follows that $\tilde{\mathcal{W}}\cap \mathcal{W}_{k-1}=\{0\}$ whence
 the composition of $L'\circ L$ restricted to $\mathcal{W}_{M-1}^{\mathrm{ann}}$
 gives an isomorphism of the $\mathfrak{h}$-module $W^*$ with the
 $\mathfrak{h}$-module $W$. By the induction hypothesis applied to
 $\mathfrak{h}$ it follows that $\mathcal{W}$ is a one-dimensional
 trivial $\mathfrak{h}$-module. Again the representation identity
 (\ref{EqRepIdentity}) in case $T$ is odd, and the identity
 (\ref{EqRepIdentityVersioni}) for even $T$ show that then
 $\mathcal{W}_M=\mathcal{V}$ is a trivial $\mathfrak{h}$-module.
 Therefore $\mathcal{V}$ can in a canonical way be seen as a simple module of the
 one-dimensional quotient algebra $\mathfrak{g}/\mathfrak{h}$ satisfying
 the hypothesis of the Theorem. But this case has already been ruled out
 in $\mathbf{2.2}$. Hence the hypothesis $\dim \mathcal{V}\geq 2$ always
 leads to a contradiction, and this proves the Theorem.

\begin{rema} {\rm The proof of Theorem \ref{DL} shows that this Theorem is still true in
the case when $\mathbb{K}$ is algebraically closed with characteristic $p\not= 2$.}
\end{rema}

\begin{coro}\label{CDL} Let $\mathfrak{g}$ be a solvable Lie superalgebra and
$\mathcal{V}$
be a non-zero finite-dimensional ${\mathfrak{g}}$-module  which admits a non-degenerate,
even,
supersymmetric and $\mathfrak{g}$-invariant bilinear form $B$.
If $\mbox{\rm dim}\mathcal{V}\geq 2$, then
there exists a non-zero isotropic
${\mathfrak{g}}$-submodule $\mathcal{W}$ of $\mathcal{V}$.
\end{coro}

\dem  If dim$\mathcal{V}\geq 2$, then, by Theorem \ref{DL}, $\mathcal{V}$ is a non-simple
${\mathfrak{g}}$-module. Consequently there exists  a
simple ${\mathfrak{g}}$-submodule $\mathcal{M}$
of $\mathcal{V}$, in particular $\mathcal{M}\not= \{0\}$,
and $\mathcal{M}\not= \mathcal{V}$.
Since $\mathcal{M}^{\perp}\cap\mathcal{M}$ is a ${\mathfrak{g}}$-submodule of
$\mathcal{M}$, then
$\mathcal{M}^{\perp}\cap\mathcal{M}= \{0\}$ or
$\mathcal{M}^{\perp}\cap\mathcal{M}= \mathcal{M}$.
It follows that either $B_{\vert_{\mathcal{M}\times \mathcal{M}}}$ is non-degenerate or
$\mathcal{M}\subseteq \mathcal{M}^{\perp}$.

If $\mathcal{M}\subseteq \mathcal{M}^{\perp}$, then $\mathcal{M}$ is a non-zero isotropic
${\mathfrak{g}}$-submodule of $\mathcal{V}$.

If $B_{\vert_{\mathcal{M}\times \mathcal{M}}}$ is non-degenerate, then, by
Theorem \ref{DL}, dim$\mathcal{M}=1$ and $\mathcal{M}$ is a trivial ${\mathfrak{g}}$-module.
Moreover, $\mathcal{V}= \mathcal{M}\oplus \mathcal{M}^{\perp}$.
Consider now a  simple ${\mathfrak{g}}$-submodule $\mathcal{M}'$ of $\mathcal{M}^{\perp}$.
If $\mathcal{M}'$
is isotropic, then the proof is finished. If $\mathcal{M}'$ is not isotropic, then
$B_{\vert_{\mathcal{M}'\times \mathcal{M}'}}$ is non-degenerate. It follows by
Theorem \ref{DL}, that $\mathcal{M}'$ is the one-dimensional trivial ${\mathfrak{g}}$-module.
Consequently, the ${\mathfrak{g}}$-module $\mathcal{M}\oplus \mathcal{M}'$ is a
$2$-dimensional trivial
nondegenerate ${\mathfrak{g}}$-module, and we can choose an isotropic
 one-dimensional graded vector sub-space
$\mathcal{W}$ of $\mathcal{M}\oplus \mathcal{M}'$. Therefore $\mathcal{W}$ is a non-zero
isotropic ${\mathfrak{g}}$-submodule of $\mathcal{V}$.

\medskip

The following Lemma is the Lie superalgebra analogue of Lemma 3.2 of
\cite{borde}:

\begin{lema}\label{martin} Let $\mathcal{V}$ be a non-zero finite dimensional
$\mathbb{Z}_2$-graded  vector space which admits a non-degenerate,
even and supersymmetric bilinear form $B$.  Let $\mathcal{L}$ a  solvable Lie
sub-superalgebra of $\mathfrak {osp}(\mathcal{V},B)$ and $\mathcal{W}$ a graded vector
subspace of $\mathcal{V}$.

If $\mathcal{W}$ is an isotropic  $\mathcal{L}$-submodule of $\mathcal{V}$, then there
 exists a graded vector subspace $\mathcal{W}_{max}$ of $\mathcal{V}$ such that:
\begin{enumerate}
\item[(i)]$ \mathcal{W}\subseteq \mathcal{W}_{max};$

\item[(ii)] $\mathcal{W}_{max}$ is an isotropic $\mathcal{L}$-submodule of $\mathcal{V};$

\item[(iii)] $\mathcal{W}_{max}$ is maximal among all isotropic graded vector sub-space
of $\mathcal{V};$

\item[(iv)] $\mbox{\rm dim}\mathcal{W}_{max}= [{n\over 2}]$ (i.e. the integer part of ${n\over 2}$);

\item[(v)] If $n$ is even, $\mathcal{W}_{max}= (\mathcal{W}_{max})^{\perp};$

\item[(vi)] If $n$ is odd, $\mathcal{W}_{max}\subseteq (\mathcal{W}_{max})^{\perp}$,
$\mbox{\rm dim}(\mathcal{W}_{max})^{\perp}-\mbox{\rm dim}\mathcal{W}_{max}= 1$, and
$f((\mathcal{W}_{max})^{\perp})\subseteq \mathcal{W}_{max}$, for all $f\in  \mathcal{L}$.
\end{enumerate}
\end{lema}
\dem We will prove this Lemma by induction on the dimension of $\mathcal{V}$. It is
clear that  that the lemma is true if dim$\mathcal{V}= 1$.

Now, let us assume that the lemma is true if dim$\mathcal{V}< n$, and we shall show it
when dim$\mathcal{V}= n$, where $n\geq 2$. By the Corollary \ref{CDL}, there exists
$\mathcal{W}$ a non-zero isotropic $\mathcal{L}$-submodule of $\mathcal{V}$. If
$\mathcal{W}= \mathcal{W}^\perp$, then dim$\mathcal{W}= {n\over 2}$, it follows that
$\mathcal{W}$ is maximal among all isotropic graded vector subspaces of $\mathcal{V}$.
Consequently, $\mathcal{W}_{max}= \mathcal{W}$. If now
$\mathcal{W}\not= \mathcal{W}^\perp$,
we consider $\mathcal{V}'= \mathcal{W}^\perp/\mathcal{W}$ which is  a
$\mathbb{Z}_2$-graded  vector space  and  $\mbox{\rm dim}\mathcal{V}'< \mbox{\rm dim}\mathcal{V}= n$
 Moreover, the well defined map
 ${\bar B}: \mathcal{V}'\times \mathcal{V}'\rightarrow \mathbb{K}$ defined by
 ${\bar B}(X+\mathcal{W},Y+\mathcal{W}):= B(X,Y), \forall X, Y \in \mathcal{W}^\perp$,
 is a non-degenerate, even and supersymmetric bilinear form.
 Now, if $f \in \mathcal{L}$, then the linear map
 ${\bar f}: \mathcal{V}'\rightarrow \mathcal{V}'$
  defined by
  ${\bar f}(X+\mathcal{W}):= f(X)+\mathcal{W}, \forall X \in \mathcal{W}^\perp$, is
  well-defined. It is clear that $\mathcal{L}'= \{{\bar f}: \,f \in \mathcal{L}\}$
  is a solvable
  Lie sub-superalgebra of $\mathfrak{osp}(\mathcal{V}',{\bar B})$. It follows, by the
  induction hypothesis,  that  there exists a graded vector subspace
  $\mathcal{W}'=\{0\}_{max}$ of $\mathcal{V}'$ which verifies the six assertions
  of the lemma. Let us consider $\mathcal{W}_{max}:= S^{-1}(\mathcal{W}')$,
  where $S: \mathcal{W}^\perp \rightarrow \mathcal{W}^\perp/\mathcal{W}$ is the
  canonical surjection. Then, $\mathcal{W}_{max}$ is a graded vector subspace
  of $\mathcal{V}$ such that $ \mathcal{W}\subseteq \mathcal{W}_{max}$ and
  $\mathcal{W}'$ is isomorphic to $\mathcal{W}_{max}/\mathcal{W}$. Consequently,
  $\mbox{dim}\mathcal{W}_{max}= [{ n\over 2}]$.

Let $X,Y \in \mathcal{W}_{max}$. Since $S(X)$ and $S(Y)$ are elements of $\mathcal{W}'$,
then  $B(X,Y)= {\bar B}(S(X),S(Y))=0$. This proves that $\mathcal{W}_{max}$  is isotropic. Consequently
$\mathcal{W}_{max}$ is maximal among all isotropic graded vector subspaces of $\mathcal{V}$
 because $\mbox{dim}\mathcal{W}_{max}= [{ n\over 2}]$. Let
 $X \in \mathcal{W}_{max},\, f\in \mathcal{L}$,  we have
 $S(f(X))= {\bar f}(S(X)) \in \mathcal{W}'$ then $f(X)\in \mathcal{W}_{max}$.
 It follows that $\mathcal{W}_{max}$ is a $\mathcal{L}$-submodule of $\mathcal{V}$.

If $n$ is even it is clear that
$\mbox{dim}\mathcal{W}_{max}= \mbox{dim}(\mathcal{W}_{max})^{\perp}={n\over 2}$.
Consequently, $\mathcal{W}_{max}= \ (\mathcal{W}_{max})^{\perp}$.

If $n$ is odd, then there exists $k \in \mathbb{N}$ such that $n= 2k+1$ and
$\mbox{dim}\mathcal{W}_{max}= k$. Consequently
$\mbox{dim}(\mathcal{W}_{max})^{\perp}-\mbox{dim} \mathcal{W}_{max}= 1$.
Now, $\mathcal{W}_{max}^{\perp}/\mathcal{W}_{max}$ is an $\mathcal{L}$-module
and  the map
${\bar B}: \mathcal{W}_{max}^{\perp}/\mathcal{W}_{max}\times \mathcal{W}_{max}^{\perp}/\mathcal{W}_{max}\rightarrow \mathbb{K}$
defined by
${\bar B}(X+ \mathcal{W}_{max},Y+\mathcal{W}_{max}):= B(X,Y),
\forall X, Y \in (\mathcal{W}_{max})^{\perp}$, is a $
\mathcal{L}-$invariant non-degenerate, even and supersymmetric bilinear form. Then,
by Theorem \ref{DL}, $\mathcal{W}_{max}^{\perp}/\mathcal{W}_{max}$ is a trivial
$\mathcal{L}$-module and, therefore,
$f((\mathcal{W}_{max})^{\perp})\subseteq \mathcal{W}_{max}$,
for all $f\in \mathcal{L}$.

\begin{theor}
Let $({\mathfrak{g}},B)$ be a solvable quadratic Lie superalgebra.
Then $\mathfrak{g}$ contains
an isotropic graded ideal $\mathfrak{I}$ of dimension
$[{{\mbox{\rm dim}{\mathfrak{g}}}\over 2}]$
which  is maximal among all isotropic graded vector subspaces of $\mathfrak{g}$.
Moreover, if $\mbox{dim}{\mathfrak{g}}$ is even then $({\mathfrak{g}},B)$
is isometric to some
$T^*-$extension of the Lie superalgebra ${\mathfrak{g}}/\mathfrak{I}$.
If $\mbox{dim}{\mathfrak{g}}$
is odd then $({\mathfrak{g}},B)$ is isometric to  a non-degenerate graded ideal of
codimension one in some $T^*-$extension of the Lie superalgebra
${\mathfrak{g}}/\mathfrak{I}$.
\end{theor}

\dem The theorem is true if $[{\mathfrak{g}},{\mathfrak{g}}]=\{0\}$
(i.e. $\mathfrak{g}$ is abelian).
Now, suppose that $\mathfrak{g}$ is not abelian.

Suppose que $[{\mathfrak{g}},{\mathfrak{g}}]\not=\{0\}$.
Since $\mathfrak{g}$ is solvable,
then $[{\mathfrak{g}},{\mathfrak{g}}]\not= {\mathfrak{g}}$ and
$\mathfrak{z}({\mathfrak{g}})\not=\{0\}$
because $[{\mathfrak{g}},{\mathfrak{g}}]^{\perp}= \mathfrak{z}({\mathfrak{g}})$.
If we suppose that the
graded ideal $[{\mathfrak{g}},{\mathfrak{g}}]$ is non-degenerate then
$\mathfrak{z}([{\mathfrak{g}},{\mathfrak{g}}])\not=\{0\}$
which contradicts the fact that
$\mathfrak{z}([{\mathfrak{g}},{\mathfrak{g}}])\subseteq
{[{\mathfrak{g}},{\mathfrak{g}}]\cap \mathfrak{z}({\mathfrak{g}})}=\{0\}$.
Consequently, $[{\mathfrak{g}},{\mathfrak{g}}]$ is degenerate.
Therefore, $\mathfrak{z}({\mathfrak{g}})\cap [{\mathfrak{g}},{\mathfrak{g}}]\not=\{0\}$.
By Lemma \ref{martin},
there exists an isotropic graded ideal $\mathfrak{I}:= (\mathfrak{z}({\mathfrak{g}})\cap [{\mathfrak{g}},{\mathfrak{g}}])_{max}$
of $\mathfrak{g}$ such
that
$\mathfrak{z}({\mathfrak{g}})\cap [{\mathfrak{g}},{\mathfrak{g}}]\subset \mathfrak{I}$,
$\mathfrak{I}$
is maximal among all isotropic graded vector subspaces of ${\mathfrak{g}}$ and
$\mbox{dim}\mathfrak{I}= [{{\mbox{dim}{\mathfrak{g}}}\over 2}]$.
If the dimension of $\mathfrak{g}$ is even, then $\mathfrak{g}$ is isometric to a
$T^*-$extension of ${\mathfrak{g}}/\mathfrak{I}$.
Now, if the dimension of $\mathfrak{g}$ is
odd, it follows, by Lemma \ref{martin},
that $\mbox{dim}\mathfrak{I}^{\perp}-\mbox{dim}\mathfrak{I}= 1$
and $[{\mathfrak{g}},\mathfrak{I}^{\perp}]\subseteq \mathfrak{I}$.
Since $B$ is non-degenerate
and invariant, then $[\mathfrak{I}^{\perp},\mathfrak{I}^{\perp}]= \{0\}$.
Since $\mbox{dim}{\mathfrak{g}}_{\bar 1}$ is even,
then $\mbox{dim}\mathfrak{I}_{\bar 1}=
{{\mbox{dim}{\mathfrak{g}}_{\bar 1}}\over 2},\, \mbox{dim}\mathfrak{I}_{\bar 0}=
[{{\mbox{dim}{\mathfrak{g}}_{\bar 0}}\over 2}]$ and
$\mbox{dim}(\mathfrak{I}_{\bar 0})^{\perp}\cap {\mathfrak{g}}_{\bar 0}
-\mbox{dim}\mathfrak{I}_{\bar 0}= 1$.
It follows that  $(\mathfrak{I}_{\bar 0})^{\perp}\cap {\mathfrak{g}}_{\bar 0}$ is not
an isotropic vector subspace of ${\mathfrak{g}}_{\bar 0}$. Consequently, there
exists $x \in  (\mathfrak{I}_{\bar 0})^{\perp}\cap {\mathfrak{g}}_{\bar 0}$ such that
$B(x,x)\not= 0$. The fact that $\mathbb{K}$ is algebraically closed implies
that there exists $\alpha \in \mathbb{K}$ such that $B(\alpha x,\alpha x)= -1$.

Let us consider the one-dimensional  Lie algebra $\mathcal{E}= \mathbb{K}e$  with its
invariant scalar product $q$ defined by $q(e,e):= 1$, and the orthogonal direct
sum $(\mathcal{A}= {\mathfrak{g}}\oplus \mathcal{E}, T:= B\perp q)$
which is a quadratic Lie
superalgebra such that
$\mathcal{A}_{\bar 0} = {\mathfrak{g}}_{\bar 0}\oplus \mathcal{E}$
and $\mathcal{A}_{\bar 1}= {\mathfrak{g}}_{\bar 1}$. Next we  consider
$f:= e + \alpha X \in \mathcal{A}_{\bar 0}$ and
$\mathcal{H}= \mathfrak{I}\oplus \mathbb{K}f$.
It is obvious that $\mathcal{H}$ is an isotropic graded ideal of $\mathcal{A}$
such that $\mbox{dim}\mathcal{H}= {{\mbox{dim}\mathcal{A}}\over 2}$. It follows
that $\mathcal{A}$ is isometric to a $T^*-$extension of the Lie superalgebra
$\mathcal{A}/\mathcal{H}$.
Let $\Phi: \mathcal{A} \rightarrow {\mathfrak{g}}/\mathfrak{I}$
be the linear map defined by:
$\Phi(x+\lambda e):= (x-\lambda\alpha X)+\mathfrak{I},\,
\forall (x,\lambda)  \in{\mathfrak{g}}\times\mathbb{K}$.
It is a surjective morphism of  Lie
superalgebras such that  $\mbox{Ker}\Phi= \mathcal{H}$. Consequently,
the Lie superalgebras ${\mathfrak{g}}/\mathfrak{I}$ and $\mathcal{A}/\mathcal{H}$
are isomorphic. We conclude that  $\mathcal{A}$ is isometric to a $T^*-$extension
of the Lie superalgebra ${\mathfrak{g}}/\mathfrak{I}$, and $\mathfrak{g}$
is isometric to a non-degenerate
graded ideal ${\mathfrak{g}}\oplus \{0\}$ of codimension $1$ of $\mathcal{A}$.


\end{document}